\documentclass{ees2023}
 \usepackage{graphicx}
 \usepackage{hyperref}
 \usepackage[]{natbib}  
 \usepackage{epstopdf}
 \usepackage{color}
 
 \def\BibTeX{{\rm B\kern-.05em{\sc i\kern-.025em b}\kern-.08em
    T\kern-.1667em\lower.7ex\hbox{E}\kern-.125emX}}
 \bibpunct{(}{)}{;}{a}{}{,}  

 \newcommand{\blue}{\textcolor{blue}}

 \newcommand{\teff}{\ensuremath{T_{\rm eff}}}
 \newcommand{\logg}{\ensuremath{\log g}}
 \newcommand{\feh}{\ensuremath{{\rm [Fe/H]}}}
 
 \newcommand{\ebmr}{\ensuremath{E(G_{\rm BP} - G_{\rm RP})}}
 
 \newcommand{\gmag}{\ensuremath{G}}
 
 \newcommand{\msol}{\ensuremath{M_{\odot}}}
 \newcommand{\lum}{\ensuremath{L_{\star}}}
\newcommand{\creditgold}{\href{https://www.aanda.org/articles/aa/full_html/2023/06/aa43688-22/aa43688-22.html}{\blue{Gaia Collaboration, Creevey et al. 2023, A\&A, 674, A39}}}
\newcommand{\creditapsisone}{\href{https://www.aanda.org/articles/aa/full_html/2023/06/aa43688-22/aa43688-22.html}{\blue{Creevey et al. 2023, A\&A, 674, A26}}}
\newcommand{\creditapsistwo}{\href{https://www.aanda.org/articles/aa/full_html/2023/06/aa43919-22/aa43919-22.html}{\blue{Fouesneau et al. 2023, A\&A, 674, A28}}}
\newcommand{\creditfrancesca}{\href{https://www.aanda.org/articles/aa/full_html/2023/06/aa43680-22/aa43680-22.html}{\blue{De Angeli et al. 2023, A\&A, 674, A2}}}

\newcommand{\creditpaoladrtwo}{\href{https://www.aanda.org/articles/aa/full_html/2018/08/aa32836-18/aa32836-18.html}{\blue{Sartoretti et al. 2018, A\&A, 616, A6}}}
\newcommand{\creditrene}{\href{https://www.aanda.org/articles/aa/full_html/2023/06/aa43462-22/aa43462-22.html}{\blue{Andrae et al. 2023, A\&A, 674, A27}}}
\newcommand{\creditarb}{\href{https://www.aanda.org/articles/aa/full_html/2023/06/aa43750-22/aa43750-22.html}{\blue{Recio-Blanco et al. 2023, A\&A, 674, A29}}}

\begin{document}
%


\title{Stellar Physics Across the HR Diagram with Gaia}

\author{O.~L. Creevey}
\address{Universit\'e C\^ote d'Azur, Observatoire de la C\^ote d'Azur, CNRS, Laboratoire Lagrange, Bd de l'Observatoire, CS 34229, 06304 Nice cedex 4, France}
\email{orlagh.creevey@oca.eu}

\phantomsection
\addcontentsline{toc}{part}{STELLAR PHYSICS ACROSS THE HR DIAGRAM WITH GAIA}



\maketitle


\begin{abstract}
Gaia Data Release 3 (GDR3) contains a wealth of information to advance our knowledge of stellar physics.  
In these lecture notes we introduce the data products from GDR3 that can be exploited by the stellar physics community.    
Then we visit different regions of the HR diagram, discuss the open scientific questions, and describe how GDR3 can help advance this particular topic.  
Specific regions include hot OB and A type stars,  FGK main sequence, giants, and variable sources,   low mass stars, and ultra-cool dwarfs.   Examples of scientific exploitation are also provided.   
These lecture notes are accompanied by a 3-hour lecture presentation and a 3-hour practical session that are publicly available on the website of the Ecole Evry Schatzman 2023: Stellar physics with Gaia, \url{https://ees2023.sciencesconf.org/}, see {\it Lectures} and {\it Hands-on Work}. 
\end{abstract}



\section{Introduction to Gaia}\label{sec:introduction}

The Gaia mission \citep{gaiamission} was launched just over 10 years ago on the 19 December 2013.  After its journey to Lagrange point L2 and its commissioning stage, it began science operations in July 2014.  Today (January 2024) it has been collecting scientific data for 9 and a half years.  Its mission extension has been approved until approximately the end of 2024 to May 2025, when the micropropulsion gas will have reached its lower limit to enable science operations.  

Gaia was designed to perform a full sky survey of about 1--2 billion of the brightest objects in visible light.  It targets a {\it G}-band range spanning approximately 3 -- 21 mag, similar to V-band but extending from 325 -- 1050 nm \citep{montegriffo23a}.   These brightest targets represent an approximate 1\% of the content of our Milky Way.  Gaia's instruments enable the collection of three main data products: astrometry, spectrophotometry, and spectroscopy.  These together allow the mission to fulfill its scientific objectives of understanding the formation and evolution of our Galaxy by measuring the accurate positions, motions, and intrinsic properties of its observed stars.  Through these measurements, Gaia is also a space mission aimed at improving our knowledge of stellar physics.   But Gaia's science does not just stop at stars and the Galaxy.  As it measures all objects spanning 3 -- 21 magnitude, it contributes significantly to the understanding of our own Solar System through the measurements of its asteroids and other minor bodies, to the understanding of sub-stellar objects such as brown dwarfs and exoplanets, and it also measures extra-galactic sources, such as quasars and  other galaxies \citep{bailerjones23,gallucio23,dibs23,apsis3}.  
A \href{https://www.michaelperryman.co.uk/gaia-essays}{\blue{series of two-page essays produced by M. Perryman}} provides an overview of the science legacy.

Gaia rotates about its own axis every 6 hours, and as it is inclined at about 45$^\circ$ relative the direction of the Sun, it also precesses.  It also follows the orbit of the Earth around the Sun.  This rotation, precession and orbital movements allow Gaia to scan the whole sky.
However, not all positions on the sky are visited the same number of times, and there are  imprints of the scanning law on the quality of the data.  Over the nominal mission lifetime of 5 years, Gaia will have observed each source a median value of 70 times (70 transits), sometimes reaching as low as 5, but also reaching a high several hundred transits.  Although the end of the nominal mission has passed and it is now in its 10th year of observations, the Gaia Data Processing and Analysis Consortium (DPAC) has of today released data based only on 34 months of science operations.  This is Gaia Data Release 3 (GDR3) which took place in June 2022 \citep{gdr3contents,babusiaux23}.   These are the data that are the subject of this school, lectures and practical work, see \href{https://ees2023.sciencesconf.org/}{\blue{https://ees2023.sciencesconf.org/}}.

Gaia's measurements are time series of positions, {\it G}-band photometry, low resolution spectro-photometry from blue and red prisms (BP and RP, or together XP), and high resolution spectroscopy (R$\sim$11,000) from the Radial Velocity Spectrometer (RVS).  
During the processing by Gaia-DPAC, these are converted to (1) an astrometric solution (positions, proper motions, and parallax), derived from the time-series positions (2) mean and time-resolved {\it G}, BP, and RP integrated photometry, along with passbands and zeropoints, and (3) mean and time-resolved XP and RVS spectra.  Another part of the DPAC processing uses these basic data products to provide astrophysical analysis; (1) analysis of multiple systems, e.g. solutions to two-body systems, (2) analysis of variable objects e.g. periods and amplitudes of pulsating stars, (3) astrophysical parametrization of all sources, e.g. class probabilities, and stellar parameters, (4) solar-system analysis, e.g. asteroid and orbit characterisation, (5) morphology and redshifts of extra-galactic sources, (6) science alerts, and (7) auxiliary products such as Gaia simulations and cross-match tables.
In GDR3 all of the above-mentioned products from the astrophysical analysis was published along with 220 million mean XP spectra and 1 million RVS spectra.  GDR3 also comprises the astrometric solution and mean photometry, as was published in Gaia Early Data Release 3 in December 2020 \citep{gaiaedr3}.   All of these data products are made available through the \href{https://gea.esac.esa.int/archive/}{\blue{Gaia Archive}}\footnote{\url{https://gea.esac.esa.int/archive/}}.  Examples of accessing these data are found in the practical sessions, and further information can be found on the \href{https://www.cosmos.esa.int/web/gaia-users/archive}{\blue{Gaia Help pages}}\footnote{\url{https://www.cosmos.esa.int/web/gaia-users/archive}}.   

In this lecture, I discuss a subset of the Gaia DR3, and focus on the stellar-related data products.
 
\section{Gaia Observables}\label{sec:observables}
To fully understand how Gaia can advance our knowledge of stellar physics, it is important to understand what we consider the {\it basic} data products; the astrometry, spectrophotometry, and spectroscopy.  I will introduce each of these in the following sections, while focussing on the products that are available in the Gaia archive, and notably in the table {\tt {gaiadr3.gaia\_source}}.

\subsection{Astrometry}
The most basic information that we can have about the content of the visible sky is a knowledge of the number of objects and their positions. The accuracy with which we can determine positions depends evidently on the brightness of a star but primarily on the spatial resolution of the instrument used to measure the positions.   By eye we can estimate the positions of the brightest targets to roughly a few arcminutes, if we can also see a source with a known reference position.  In fact the definition of 20/20 vision is the ability to resolve to one arcsecond.  However, at night, conditions are not optimal, and the atmosphere induces aberrations, which increases this limit to a few arcminutes.    
This also means that if two stars are closer than a few arcminutes, then we are probably not capable of observing, or resolving, the individual sources.  The use of binoculars can help go further allowing us to increase the resolving power by a factor typically about 10, and of course modern telescopes increase this number further.  
   
\begin{figure}
\centering{
\includegraphics[width=0.48\textwidth]{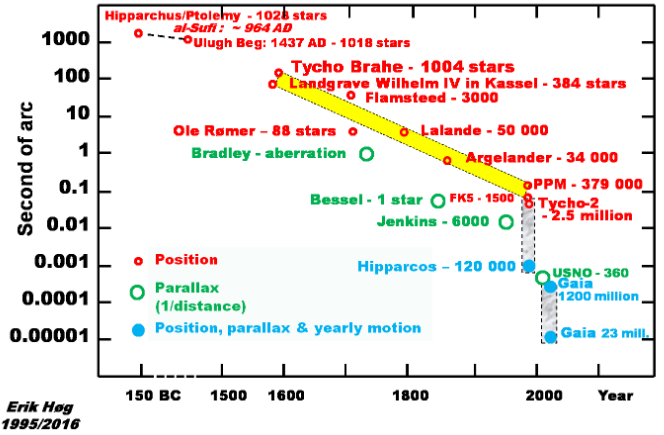} 
\includegraphics[width=0.49\textwidth]{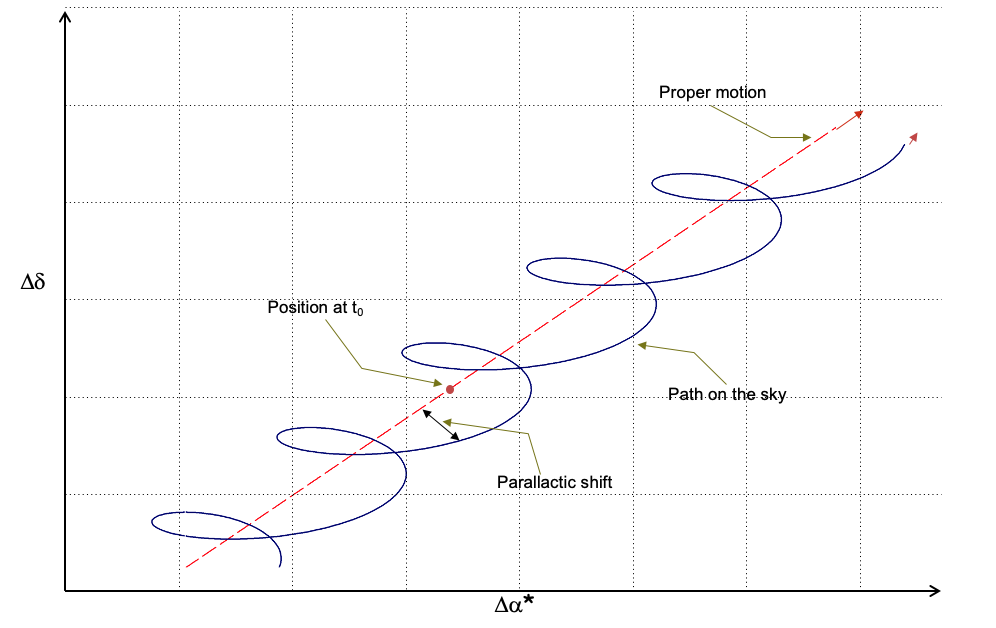}
\caption{{\sl Left:} The precision in star positions in the sky for catalogues spanning the last 2000 years. Credit: \href{https://ui.adsabs.harvard.edu/abs/2024arXiv240210996H/abstract}{\blue{H{\o}g, E. 2024, arXiv:2402.10996}}.
{\sl Right:} The apparent motion of a star on the sky as we observe it from different positions and over time (blue).  The red dashed line shows the proper motion of the star, and the black arrow indicates the parallactic shift as the observer's position moves around the Sun.  Approximately five years of measurements is illustrated.  
Acknowledgement: Figure kindly provided by Fran\c{c}ois Mignard.}
\label{fig:positionaccuracy}}
\end{figure}

One of the first catalogues of stars to be produced was by Hipparchos of Rhodes in approximately 130 BC.  His catalogue comprised about 850 stars with a typical precision on the order of a degree (at that time there was no known reference position).  Roughly 1700 years later, Tycho Brahe measured about 1000 stars with a roughly 1 arcminute precision.   As we advanced in our knowledge of the sky, measurement techniques, and instruments, the numbers of stars and accuracy of their positions increased significantly to roughly one arcsecond until modern techniques and surveys appeared, see Fig.~\ref{fig:positionaccuracy} left panel which illustrates the increasing precision of astrometric measurements over time, along with appearance of reference catalogues \citep{hog17}.

The increased precision of positions also allowed the measurements of the relative positions of stars with respect to their {\it fixed} background.  If we observe a star at two distinct moments, the star will appear to move with respect to the fixed background.  How much it moves will depend on how far away the star is and the difference in angle from which we measure it.  
This apparent movement in the sky is called a parallax, and it allows one to estimate the star's distance by using trigonometry if we know the distance between the two points when the measurements were made.   The use of the Earth's orbital motion around the Sun allows us to do just this, by measuring the position of the star relative to the background stars at a separation of 6 months when the Earth is exactly 2 astronomical units (AU) away from its first position. 
A stellar parallax was first measured by Bessel for 61 Cygni in 1838. He deduced a parallax of $314 \pm 20$ milli arcseconds (mas) which provided the first distance estimate of roughly 10 light years.   Knowing the distance to objects allows one to estimate their true luminosities and radii, two  of the fundamental parameters of stars which are a product of stellar structure and evolution.   

The first catalogue to allow mas precision on the parallaxes of stars was the Hipparcos mission \citep{hipparcos97}.  It measured the parallaxes of 120\,000 stars up to a visible magnitude of roughly 10.  This catalogue remained the reference catalogue until Gaia Data Release 2 in 2018 for most of these stars, except for the brightest where Gaia needs its full nominal mission data to better understand the systematics at these low magnitudes.

\subsubsection{An astrometric solution}
Given a set of position measurements of a star at specific time points, one can trace the apparent motion of a star with reference to fixed points in the background, see Fig.~\ref{fig:positionaccuracy} right panel.  For a single star, the parameters that are solved are (1) the mean position in right ascension (ra) and declination (dec) at a reference time denoted $t_0$, usually taken to be the central time of the span of the measurements, (2) the mean motion of the star on the projected sky background, referred to the proper motion in ra* and dec, and (3) the parallax, which is the radius of the circle that is traced per year of observation.  These five parameters provide an astrometric solution for a single star.    Due to telescope optics, the colour of the star will also play a role in the measured positions of the star, so that one must also consider this parameter.  In Gaia DR3 the {\tt pseudocolor} or {\tt nueff\_measured\_in\_astrometry} are these parameters.  In the first case the colour is fitted (6-parameter solution) while in the second it is fixed (5-parameter solution).  
The number and quality of the measurements covering the apparent motion on the sky over the 34 months determines the uncertainties and correlations in the solved parameters.
The purposes of these lecture notes is not to go in detail on how Gaia produces astrometric solutions, readers are therefore referred to e.g. \cite{2021A&A...649A...2L}, and \cite{brown21} provides an annual review while describing the astrometric problem, the solution and some science highlights using Gaia data.  Here we just focus on understanding the contents of the archive through a qualitative understanding of the data.

\subsubsection{Archive parameters}
The astrometric solution is what gives rise to each source identification.  The main identifier in most of the archive tables is the {\tt source\_id} which is a set of between 16 and 20 numbers used to encode information concerning the position of the source in the sky, the processing center used to produce the solution, and a running ID.  For example, if one would like to obtain the level 12 HEALPix identification, one should divide the source ID by 34359738368.  It is important to note that with each processing cycle, that is with each main Gaia Data Release, the source ID can change.  If one would like to crossmatch the source ID with another catalogue, one should use the designation `Gaia DR3 {\tt source\_id}', e.g. `Gaia DR3 1872046574983497216'.

Each source ID has an associated ra and dec, along with proper motions and a parallax if the solution has at least five parameters, i.e. {\tt astrometric\_parameters\_solved} = 31 or 95.  If this number is 3, then it indicates that the astrometric solution has only fitted two parameters (ra, dec).
Along with the main parameters, there are also several fields published pertaining to their uncertainties and correlations, e.g. {\tt ra\_parallax\_corr}.
Other parameters related to the astrometric solution concern the model parameters, e.g. {\tt astrometric\_parameters\_solved} or {\tt nu\_eff\_used\_in\_astrometry}.
Additional information is made available concerning the quality of the model fit such as a chi-square, unit weight error.  Examples of these parameters are {\tt astrometric\_gof\_al}, {\tt astrometric\_chi\_al}, or {\tt astrometric\_excess\_noise}, where the {\tt al} refers to the {\it along-scan} direction of the analysed field, and {\tt ac} is perpendicular to this, otherwise known as {\it across-scan}.  The latter parameter indicates how close the model is to the data, assuming a single star solution.  A large value could therefore indicate a non-single source, or issues with convergence.  Of particular interest is the {\tt ruwe} or {\it re-normalised unit weight error} defined as $ruwe=\frac{\sqrt{chi/(goodobs-m)}}{f(G, G_{\rm BP} - G_{\rm RP})}$ where {\it chi} is the astrometric chi-square, {\it goodobs} is {\tt astrometric\_n\_good\_obs\_al} and m is the effective number of model parameters.  It is similar to a reduced chi-squared, but modified by a function (or renormalised) that depends on $G$ and the source's colour.  A value of 1 is an ideal value for a single star solution, and a value larger than e.g. 1.5 could indicate a binary.  See the official \href{{https://gea.esac.esa.int/archive/documentation/GDR3/Data_processing/chap_cu3ast/}}{\blue{online documentation}} for further details.

Additional information from the astrometric processing is also made available, such as the number of observations used in the solution, the number of peaks that are found in the images, or if the source is a {\tt duplicated\_source}.  This latter could indicate cross-match or processing issues, or duplicity of a source, knowing that Gaia catalogue has a spatial resolution of 0.18 arcsec in GDR3 (compare to 0.4 arcsec in GDR2).

\subsubsection{Astrometric biases}
As detailed in \cite{2021A&A...649A...4L}, biases are present in the parallaxes of the astrometric solution.  In particular for stars, these biases have been shown to be a function of magnitude, colour and position in the sky, see e.g. Fig.~19 from the above reference.
A parallax zeropoint bias correction, $\pi_{\rm cor}$, has been proposed by the authors, to be used as judged relevant by a user of the archive.  This bias takes the form of a correction where $\pi_{\rm cor} = \pi_{\rm GDR3} - Z_N$ where $N$ is 5 or 6, referencing the number of parameters solved, and $Z_N$ is a function of the three above mentioned parameters.   A python function has been made available to the public by one of the authors, called {\tt gaiadr3\_zeropoint}, which returns the function $Z_N$.  The code can be found on \url{https://gitlab.com/icc-ub/public/gaiadr3\_zeropoint}.

\subsection{Spectrophotometry}
\begin{figure}
\includegraphics[width=0.9\textwidth]{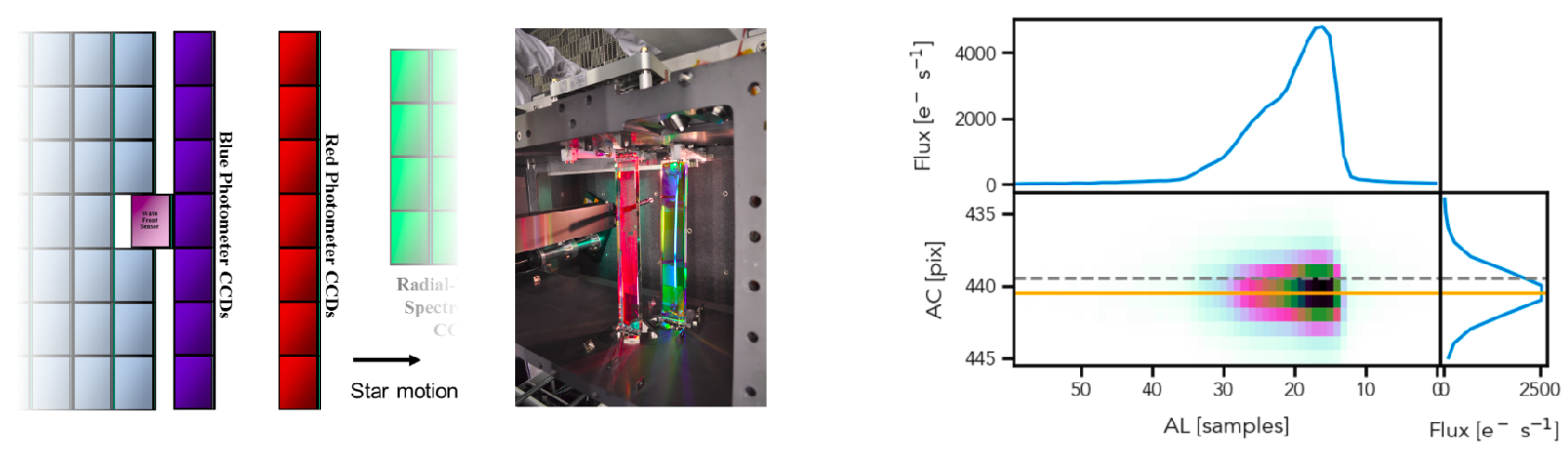}
\caption{Composite image provided by F. De Angeli illustrating the BP and RP prisms on board Gaia (center), the position of the prism CCDs in the focal plane (left) and an example of the BP/RP images analysed by the DPAC processing (right).  Credit: \creditfrancesca.} 
\label{fig:spectrophotometry}
\end{figure}

Two prism spectra are on-board Gaia which together provide low resolution spectra (resolution between 30 -- 90, \citealt{deangeli23}) spanning 330 nm -- 680 nm (BP) and 620 nm -- 1050 nm (RP), see Fig.~\ref{fig:spectrophotometry}.  The processing of these prism spectra provide two main data products (both mean and time series): (1) low resolution spectra, and (2) integrated photometry, $G_{\rm BP}$ and $G_{\rm RP}$.  The $G$ band photometry is also produced by the same processing unit by exploiting the astrometric CCDs.  Details of the data reduction, processing and calibrations are given in \cite{riello21} for the photometry, \cite{deangeli23} for the spectral processing, and \cite{montegriffo23a} for the external calibration of the spectra.
    
\subsubsection{Use in DPAC}
\begin{figure}
\includegraphics[width=0.47\textwidth]{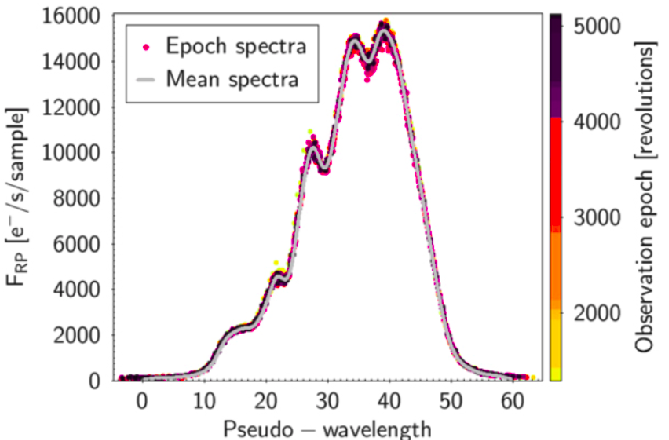}
\includegraphics[width=0.53\textwidth]{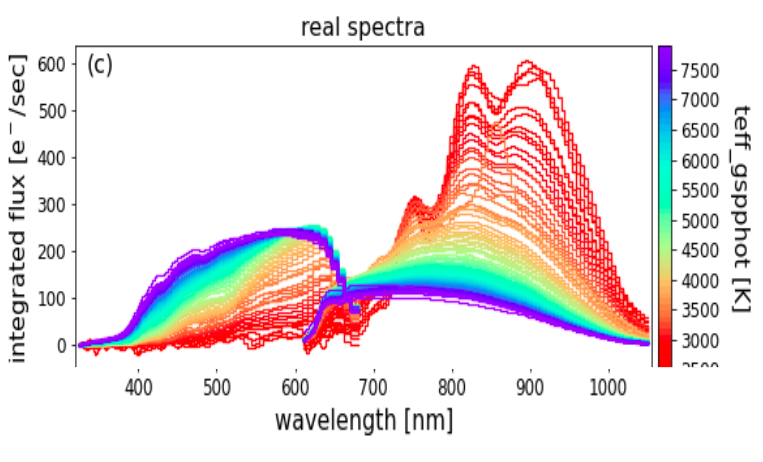}
\caption{{\sl Left:} 
RP epoch spectra for one source with the mean spectrum overlaid.  The colour-code indicates at which time the data were measured.  The y-axis shows pseudo-wavelength which can be converted to wavelength using a dispersion law. Credit: \creditfrancesca.
{\sl Right:} 
Examples of sampled mean spectra of various sources with low extinction for different \teff.   Credit: \creditapsisone.}
\label{fig:rp-meanepoch}
\end{figure}

All subsystems within DPAC use a subset of these data products.  Analysis of source variability (pulsating stars, eclipsing binaries, ...) is performed with the time series products, while source characterisation (source type, effective temperature, mass, ...) is performed with the mean data products.     These spectra are produced from the individual time series and use a Hermite polynomial representation.   Fig.~\ref{fig:rp-meanepoch} left panel illustrates a set of RP epoch spectra for an individual source, represented by the coloured dots, where the colour is indicative of its time stamp.  The mean spectrum is overlaid, giving an accurate representation of this source's RP spectrum.  The x-axis is pseudo-wavelength which can be converted into a physical wavelength by using a dispersion law.  The shape of this spectrum is due to the instrument.  These internally calibrated spectra can be converted to a true spectral energy distribution having knowledge of the instrument model.  
The internally calibrated mean spectra are used for producing astrophysical parameters and for the analysis of variable sources.  An example is given on the right panel of Fig.~\ref{fig:rp-meanepoch}, where the BP and RP spectra of stars with low extinction and different effective temperatures \teff\ are shown.

\subsubsection{Archive parameters}
220 million internally calibrated mean-spectra  have been made available in GDR3.  These are mostly for sources with $G < 17.6$ but many subsets of fainter sources have also been published.  100 million sampled mean spectra have also been published.  These data products are found through the Gaia archive datalink.    
A python tool {\tt GaiaXPy} to exploit these data has been made available at \url{https://gaia-dpci.github.io/GaiaXPy-website/}, and an example of its use is given in the \href{https://gricad-gitlab.univ-grenoble-alpes.fr/ipag-public/gaia/ees2023}{practical course}.  
In {\tt gaiadr3.gaia\_source} there are several fields related to the BP/RP processing, such as the number of photometric observations used for the processing of each prism e.g. {\tt phot\_rp\_n\_obs}, and the number of contaminated or blended transits e.g. {\tt phot\_bp\_n\_blended\_transits}, and one may want to use these values to judge the quality of the spectra or an indication if the source may be contaminated by a nearby star.

Concerning the integrated photometry, derived colours, such as {\tt bp\_rp}, and several statistical parameters, such as the mean photometry (e.g. {\tt phot\_bp\_mean\_flux}) and error
are made available in the catalogue.
Passbands and zeropoints are also associated with the photometry and these are available at \url{https://www.cosmos.esa.int/web/gaia/edr3-passbands}. Another interesting parameter is the excess factor which gives the ratio of the sum of the BP and RP integrated flux compared to the G integrated flux.  A value deviating from the norm could indicate inconsistencies of the fluxes, in particular due to crowding issues, see \cite{riello21} their Sect.~9.4 for details.  In GDR3 the precision in the photometry is 0.25 mmag at $G<13$, 1 mmag at $G<17$ and 5 mmag at $G<20$.  These values increase by a factor of 5 to 20 for $G_{\rm BP}$ and $G_{\rm RP}$.

Time series photometry for variable objects has also been published in GDR3, see lectures and practical work in this school by \cite{laurent}.  These sources are listed in the {\tt gaiadr3.vari\_summary} table, and $G$, $G_{\rm BP}$ and $G_{\rm RP}$ time series for these sources  are available through datalink.
In addition to the mean and time series photometry, a $G$-band bolometric correction tool is available under Gaia DR3 software tools and can be accessed directly here: \url{https://www.cosmos.esa.int/web/gaia/dr3-bolometric-correction-tool}, produced by the group providing astrophysical parameters, see \cite{apsis1}.

\subsection{Spectroscopy}
\begin{figure}
\includegraphics[width=0.6\textwidth]{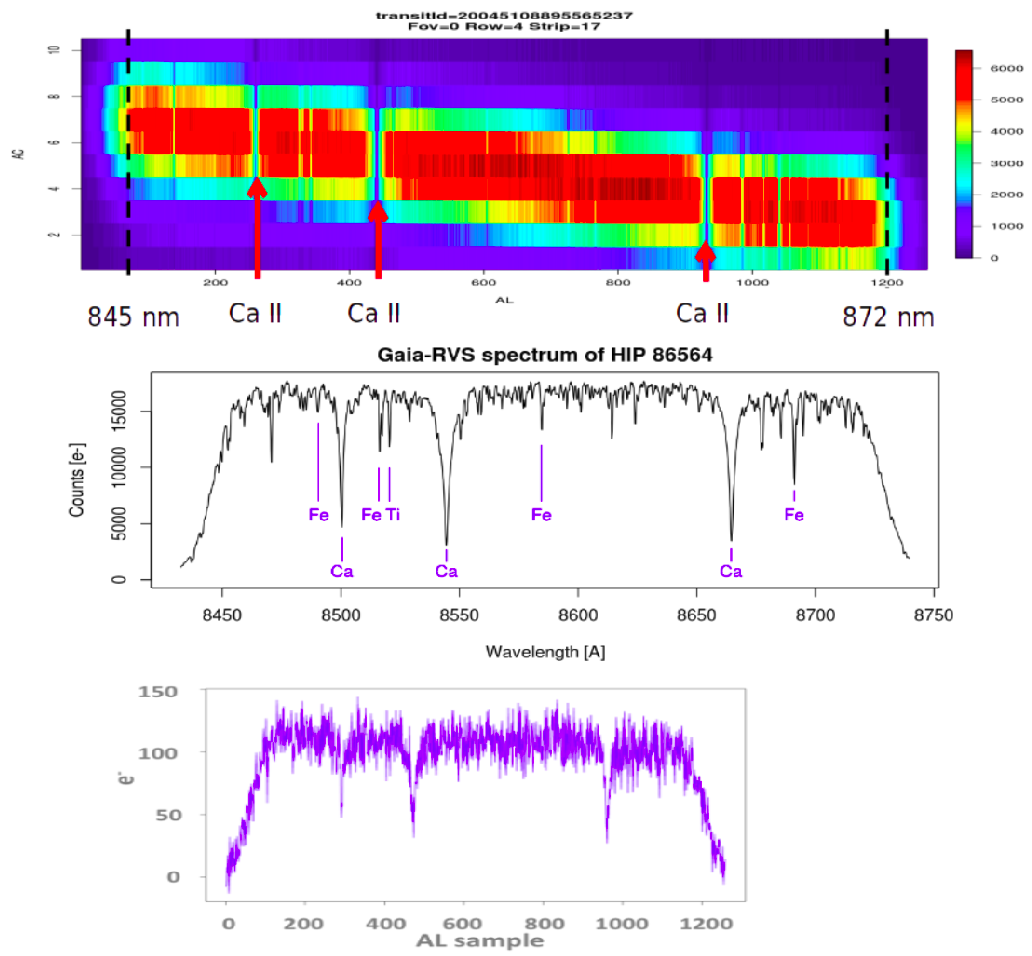}
\includegraphics[width=0.35\textwidth]{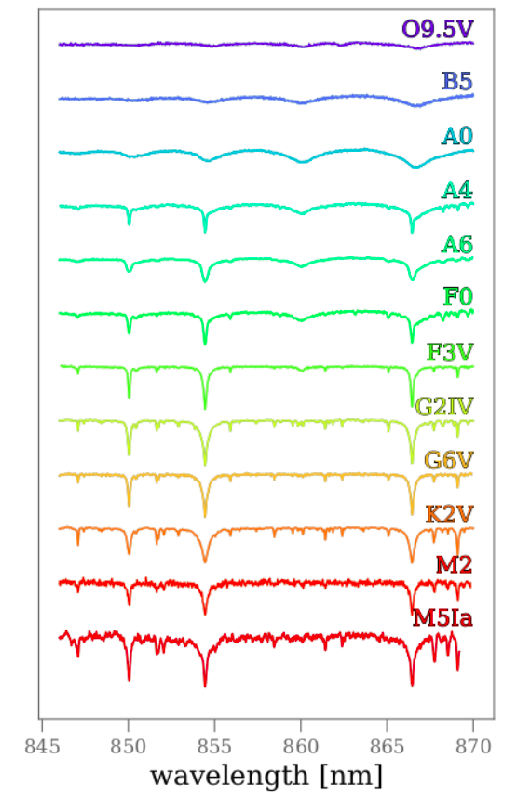}
\caption{{\sl Left:} Top panel shows the RVS pixel spectra spanning 845 nm to 872 nm with the three Ca lines (the Calcium triplet) indicated by the red arrows.   The middle panel shows an example of a 1D RVS spectrum derived from the processing for a high signal-to-noise spectrum, while the lower panel illustrates a noisier spectrum. Credits: Top panel: \creditpaoladrtwo.  Middle panel: \href{https://www.cosmos.esa.int/web/gaia/iow_20141124}{\blue{ESA/Gaia/DPAC/Observatoire de Paris-Meudon/Olivier Marchal \& David Katz}}, Lower panel: Figure produced by O. Marchal.  {\sl Right:}  A spectral library showing RVS spectra for stars of spectral types 0 to M.  Credit: \creditapsistwo.}
\label{fig:rvsspectra}
\end{figure}

A third instrument on-board Gaia is the Radial Velocity Spectrometer, RVS, a spectrograph with a resolution on the order of 11\,000 spanning the 845~nm -- 872~nm range and observing sources with $G<16.2$.  Fig.~\ref{fig:rvsspectra} left panel illustrates an example of a CCD image of a spectrum (top), along with the final data product that is produced after the spectral processing (middle), see \cite{sartoretti18} for precise details on this processing.  As one can see, several prominant spectral lines can be seen, notably the Ca IR triplet at 849.8 nm, 854.2 nm and 866.2 nm.  However, this is a very high signal-to-noise-ratio (SNR) spectrum, and many of the RVS spectra have much lower SNR, as illustrated in the bottom left panel.  On the right panel, a spectral library is shown illustrating qualitatively the prominant features available for stars of different spectral types, or \teff.  One can appreciate the difficulty of deriving physical quantities for a hot star of spectral type B or O given the lack of features available in the spectrum.     

The RVS has been designed primarily to measure the radial velocity (RV) of each source, which when complemented with the astrometric solution, gives the full position and velocity motion of a star.   For stars with $G < 12$ the RVs are measured on an epoch basis and the mean RV is derived, while for fainter targets the RVS epoch spectra are first combined and the RV is subsequently measured.  The precise measurement of the RVs need {\it a priori} a precise characterisation of the star's atmospheric parameters, \teff, iron abundance, \feh, and surface gravity \logg.  It also requires knowledge of spectral broadening and the instrument's line spread function, which are determined during their processing.  Therefore, an error induced in any of these parameters, could result in an erronous RV.   Discussion and validation of the RVs are described in \cite{katz2023}.
In the case of binarity,  RVs of single (SB1) or double (SB2) lines can be measured.  These time series of RVs, along with the time series of positions and photometry, are then treated by the non-single star subsystem to provide orbital parameters and, when possible, individual or minimal masses of one or both components.  Lecture notes and practical work on non-single stars is discussed further in this school in \cite{frederic}, see also \href{https://gricad-gitlab.univ-grenoble-alpes.fr/ipag-public/gaia/ees2023}{the hands-on work with spectra}.

\subsubsection{Use in DPAC}
The RVS spectra are used primarily to derive mean RVs of single stars \citep{katz2023} and time series of multiple systems and classical pulsating stars such as RR Lyrae and Cepheids \citep{clementini23}.  The latter data products are then interpreted by the non-single star subsystem, as mentioned above, and the variability subsystem to produce pulsation properties.   The mean RVS spectra are also used for astrophysical characterisation to produce atmospheric properties of sources \citep{apsis1,apsis2,recioblanco23,yveslec}, and when possible,  chemical abundances, see the lecture notes in this school by \cite{alejandra}.

\subsubsection{Archive parameters}
In {\tt gaiadr3.gaia\_source} a set of parameters produced by the RVS processing system are published, along with a sample of approximately 1 million RVS spectra, with SNR of between 5 and 300.  The parameters comprise 34 million RVs and errors for $G_{\rm RVS} < 16$, where $G_{\rm RVS}$ comes directly from the flux in the spectra, see \cite{sartoretti2023}, and is also available as a published parameter called {\tt grvs\_mag}.
Processing and model-fitting information is also provided, such as the number of observations used {\tt rv\_nb\_transits}, number of visibility periods (a group of measurements taken in a short time span), or the method applied to process the RVs {\tt rv\_method\_used}.  Model-fitting information examples include the $\chi^2$ p-value, goodness of fit, and template atmospheric parameters used, e.g. {\tt rv\_chisq\_pvalue}, {\tt rv\_template\_teff}.
In addition to these parameters an estimate of the broadening induced on the spectral lines due to many physical processes in the atmosphere such as rotation, convection, and winds, is also provided and summarised as the {\tt vbroad} parameter.   \cite{fremat23} shows that the broadening parameter is often a good estimate of the $v \sin i$, but how well it is a good representation depends on the \teff, and how accurately the \teff\ can be measured, see Figs.~2 and 11 of that paper.

All parameters related to multiplicity, variability, or the astrophysical parameters are found in dedicated archive tables and will be discussed in more detail below and subsequent lectures.

\subsection{Accessing multi-dimensional products}
Multi-dimensional products are data products that can not be represented by a single value in an archive table, such as the RVS and XP spectra. Other products include Markov Chain Monte Carlo (MCMC) samples, see Sect.~\ref{sec:fgkstars}. 
These can be accessed after a query on the archive by clicking on the {\tt datalink} button of the query results.   They can also be accessed programmatically through scripts e.g. by using python.   Concrete examples are provided in the \href{https://gricad-gitlab.univ-grenoble-alpes.fr/ipag-public/gaia/ees2023}{practical session}.

The \href{https://gea.esac.esa.int/archive/}{\blue{single-object tab of the archive}} 
provides detailed information on individual sources, and allows a unique view of the object without having to download all of the products.  
Astrometric, photometric, radial velocity, and astrophysical parameters (as given in {\tt gaiadr3.gaia\_source}) are highlighted in a human-readable table form.  The position of the star on the sky is shown in a skymap using ESASky.  
A visualisation of all available datalink and other time series data for this source are also shown.   For users interested in knowing more about a particular source, this single-object tab is extremely useful.

\section{Stellar physics across the HR diagram}\label{sec:stellarphysics}
Given the primary data products all described in Sect.~\ref{sec:observables}, a wealth of astrophysical information on stars and other sources can be derived and used to improve our knowledge of stellar physics.   Dedicated processing modules in DPAC provide a first homogenous analysis of the primary data products and these are made available to the community.   Such examples are periods and amplitudes of pulsating stars found in the {\tt gaiadr3.vari\_*} tables, stellar parameters from the BP/RP and RVS spectra found in the {\tt gaiadr3.astrophysical\_parameters*} tables, or orbital information on multiple star systems found in the {\tt gaiadr3.nss\_*} tables, where `*' indicates the existance of multiple tables with the same root name.   
The rest of the lecture is dedicated to explaining this content of Gaia DR3 in the context of stellar physics.   This will be done by addressing different regions of the HR diagram, where I will  indicate the open questions in stellar physics and the available parameters in GDR3 that can be used to address these questions.   I will also give a few examples of scientific exploitation of GDR3 in specific stellar regimes.   Complementary dedicated lecture notes, presentations and practical work are provided in this school by \cite{alejandra,celine,frederic,laurent,yveslec}, which enter into more detail on specific topics.
Before embarking on the visit through the HR diagram, I will begin with a brief introduction to the Astrophysical Parameters Inference System (Apsis, see \citealt{apsis1}) which is the subsystem responsible for the characterisation of a half billion stars (\teff, \logg, radius, extinction, ...).

\subsection{Introduction}
\begin{figure}
\center{\includegraphics[width=0.9\textwidth]{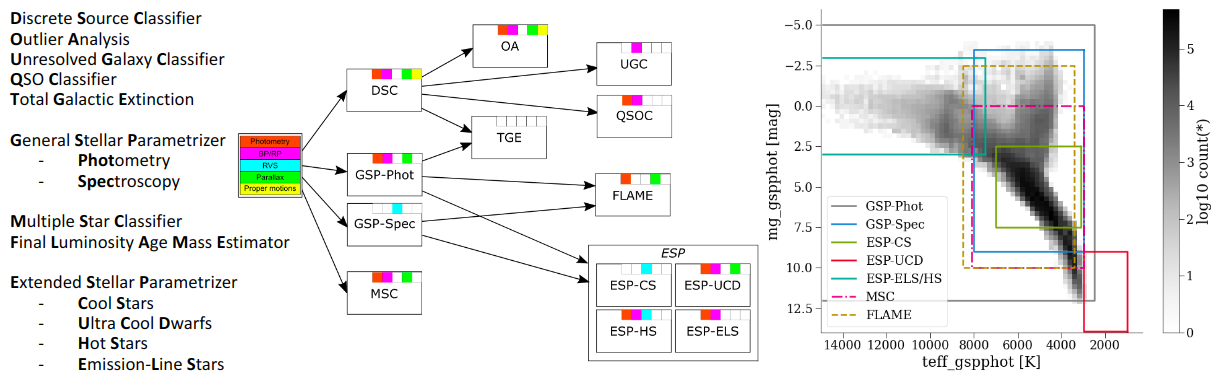}}
\caption{The Apsis system comprises several modules that derive stellar and non-stellar parametrization and classification.  Each module is listed on the left of the figure, and it is represented by a box in the functional diagram in the middle panel.  The box also indicates which data are used for the processing and the arrows show the dependencies among modules.  On the right panel a HR diagram is shown using the stellar parameters from Apsis.  The different coloured boxes indicate the region of the diagram which are analysed by the different stellar-based modules. Credit: \creditapsisone.}
\label{fig:apsisHR}
\end{figure}

Astrophysical parameters (APs) in GDR3 are produced by the Apsis analysis system in DPAC \citep{apsis1}. 
Apsis exploits the astrometry, {\it mean} BP/RP and RVS spectra, and {\it mean} photometry along with stellar models and/or training data to derive a range of AP products.  These products include probabilities and classifications (star, galaxy, spectral type,...), outlier object analysis through the use of self-organising maps, QSO and unresolved galaxy redshifts, interstellar extinction parameters such as $A_0$ (extintion parameter defined at a reference value of 547.7 nm) and a 2D total galactic extinction map, and stellar parametrization.  Among the stellar parameters, Apsis estimates 470 million spectroscopic parameters from the BP/RP spectra and 6 million from the RVS spectra (e.g. \teff, \feh), 220 million evolutionary parameters (e.g. radius, age) and H-$\alpha$ equivalent widths, 2 million activity indices and about 1 million sources with many chemical abundances.    These are produced by specific modules within the Apsis system that are designed to target a specific region of the HR diagram or use distinct data sets.   Fig.~\ref{fig:apsisHR} summarises the names of the Apsis modules on the left, the input data and dependencies for each module in the middle, while the figure on the right highlights the different regions of the HR diagram that are analysed by the stellar-based modules.  In GDR3 the parameters from the stellar modules are found in the {\tt gaiadr3.astrophysical\_parameters} and {\tt gaiadr3.astrophysical\_parameters\_supp} tables, with a very small subset copied to {\tt gaiadr3.gaia\_source}.  The field names have their modules encoded in them, so the user can trace the methods and models used to derive the parameter, see \href{https://gea.esac.esa.int/archive/documentation/GDR3/Data_analysis/chap_cu8par/}{\blue{Chapter 11 of the Gaia DR3 online documentation}}. 

\begin{figure}
\center{\includegraphics[width=0.52\textwidth]{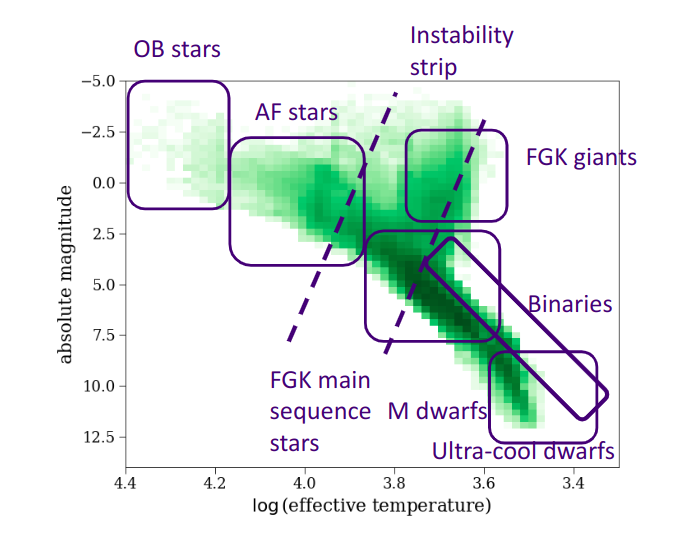}
\includegraphics[width=0.40\textwidth]{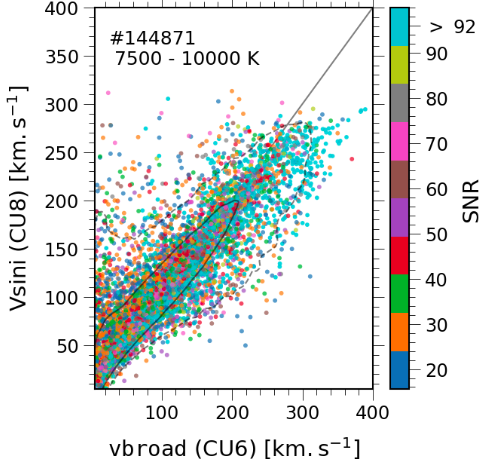}}
\caption{{\sl Left:} Different regions of the HR diagram presented in these lectures.
{\sl Right:} A comparison between two independent determinations of line broadening; x-axis {\tt vbroad} and y-axis {\tt vsini\_esphs}.  Credit:
\href{https://gea.esac.esa.int/archive/documentation/GDR3/Data_analysis/chap\_cu8par/sec\_cu8par\_validation/ssec\_cu8par\_qa\_atmospheric-aps.html\#Ch11.F82}{\blue{Ulla, A. et al., 2022, Gaia DR3 Online Documentation}}.}
\label{fig:stph-hrdiag}
\end{figure}

There are certain regions of the HR diagram that are discussed in detail in the lecture notes by \cite{alejandra,celine,frederic,laurent,yveslec}.  These are pulsating stars and binaries which exploit the time series data, the hottest and coolest stars in the HR diagram, and chemical abundances.  These will be mentioned in these lecture notes, but the reader will be referred to those lectures for details.  Fig.~\ref{fig:stph-hrdiag} illustrates a HR diagram with the GSP-Phot parameters from Apsis, with the colour-code indicating relative number density.  We have split the diagram into several regions and the lecture notes that follow will address each of these regions individually.

The Apsis module GSP-Phot \citep{andrae23} analyses the mean BP/RP spectra of each source using four libraries of stellar atmospheres with a \teff\ range between 2\,500 and 50\,000 K and fits the atmospheric parameters \teff, \logg, \feh, and $A_0$ and distance assuming a single star.   The assumptions laid out here explain the following few attributes in Fig.~\ref{fig:stph-hrdiag}:  (1) there are no white dwarfs because  these models are not included, and therefore are assigned incorrect parameters, resulting in poor posterior distribution functions (pdf); (2) there are no dwarfs below 2\,500 K, although we have highlighted the ultra-cool dwarfs because these are analysed by a different stellar module and these will be discussed further; (3) the binary sequence is not well represented, although one could use the diagram to estimate which sources are binaries; (4) some patchiness is seen and this is partially due to parts of the diagram having a relatively low number of stars, but also due to the boundaries in the model grids mostly seen at 6\,000 K and 15\,000 K.    Concerning the white dwarfs, a user can exploit directly the photometry and astrometry to easily identify those sources, see e.g. Figs.~1 or 12 from \cite{carine18}.

To complete the information about GSP-Phot, as it will be mentioned several times in the following sections, all sources are analysed using four libraries of stellar atmospheres.  These are OB, A-star, PHOENIX, and MARCS.   The results for all sources for all libraries are found in the {\tt gaiadr3.astrophysical\_parameters\_supp} table.  However, GSP-Phot also chooses the most appropriate library for each source based on convergence criteria.  This best library for each source is given in the {\tt gaiadr3.astrophysical\_parameters} table and is indicated in the {\tt libname\_gspphot} field, and a subset of those parameters are also found in {\tt gaiadr3.gaia\_source}.

\subsection{Hot stars}\label{sec:hot}
\subsubsection{OB spectral types}\label{sec:ob} 
O and B spectral-type stars are intermediate to massive stars. Due to their large masses and therefore very high central pressures, and thus temperatures, they consume their nuclear fuel at a very efficient rate, and thus evolve very rapidly.  Due to the high central temperatures, they also go through several nuclear-burning phases and produce the heaviest elements.  They are therefore key contributors to the enrichment of the interstellar medium through supernova.     As they remain main sequence stars for such a short time they do not move far from their birth place, and this fact makes them excellent probes of the structure and dynamics of star forming regions and spiral arms, see e.g. \cite{drimmel23,gold}.   

In stellar physics OB stars contribute to our understanding of: 
\begin{itemize}
\item  mass-loss due to winds from the important contribution of radiation pressure,
\item observations of their photospheres and stellar oscillations allow us to test opacities as produced by laboratory experiments, 
\item   as these objects have many He lines, they can provide observational measures of He abundances and can help to constrain chemical enrichment laws, as well as constrain chemical transport mechanisms,
\item the identification of the main sequence stars allows one to put important constraints on their ages (between 10 million -- 200 million years), in particular if there is a companion or if the star is a member of a cluster, 
\item the more evolved OB stars become blue horizontal giant branch stars, are progenitors of supernova, and later become neutro`B'n stars or black holes, and are potential contributors to detectable gravitational waves.

\end{itemize}

In GDR3, one of the key products for OB stars is their identification, and then initial characterisation.  Within Apsis, the ESP-ELS (extended stellar parametrizer -- emission line stars) module provides firstly a spectral typing of all sources up to $G < 17.65$, by classifying them  into the broad categories of `O',`B', `A', `F', etc (in the archive the parameter is labelled {\tt spectraltype\_esphs}).   Once the star has been identified, it is then tested if it is an emission line star, assigned a probability of being a particular emission-type, e.g. {\tt classprob\_espels\_wcstar},  and then finally classified according to its emission as a Be star (beStar), Herbig Ae/Be star (HerbigStar), T Tauri star (TTauri), active M dwarf star (RedDwarfEmStar),
Wolf-Rayet WC (wC) or WN (wN), or a planetary nebula (PlanetaryNebula) e.g. {\tt classlabel\_espels = wC}.
The stars that have been identified as O, B, or A, are further analysed by the ESP-HS (hot star) module, which exploits the mean BP/RP and RVS spectra, to produce \teff, \logg, $A_0$, \ebmr, and $v \sin i$ if the RVS spectrum is available, with the assumption of solar metallicity.  
These parameters are also found in the {\tt gaiadr3.astrophysical\_parameters} table and are called {\tt teff\_esphs}, {\tt logg\_esphs} and {\tt vsini\_esphs}.   If the RVS spectrum is not available, then $v \sin i$ is not derived.   No specific treatment of rotation is included in the models that are used to produce the simulations of the low-resolution BP and RP spectra.

A second set of OB characterisation is provided by GSP-Phot, which includes its \teff, \logg, \feh, distance, $A_0$ and \ebmr, by using the OB star library, and found in the {\tt gaiadr3.astrophysical\_parameters\_supp} table.  All sources have a solution with the OB library, so one should take care in selecting the sources.  One could alternatively use directly the result in the {\tt gaiadr3.astrophysical\_parameters} table, where the {\tt libname\_gspphot} is `OB' i.e. the preferred solution for this star is found to be `OB' and not `A', `MARCS' or `PHOENIX'. 
If results from both ESP-HS and GSP-Phot are available, the user should prioritise the former where both RVS and BP/RP spectra are used, because the models have been more finely tuned to the data focussing on hot stars.    Further information on hot stars is given in the lecture by \cite{yveslec}. 

Finally, just to be complete, although this is applicable to all sources, kinematic information such as proper motions, radial velocity and positions for these stars are given in {\tt gaiadr3.gaia\_source}.

\subsubsection{AF spectral types}\label{sec:af}
Spectral-type A and F stars have masses between approximately 1.4 -- 2.1 \msol, and \teff\ in the range of 7\,600 -- 10\,000 K.  Their spectra are dominated by H lines.   Some famous examples are Vega and Altair, and a well known planet-host $\beta$ Pic.   They are in general rapid rotators but their rotational velocity changes rapidly with sub spectral type (or \teff) and more slowly with stellar evolution.   Being intermediate mass stars, their typical main sequence lifetimes are on the order of about 1 billion years, and this makes them interesting probes for stellar ages, in particular in clusters.    Many of these stars are also of variable type known as $\delta$ Scuti or $\gamma$ Dor, or hybrid-mode, and show large amplitude pulsations that can usually be detected from ground-based observatories.   For this reason, these are some of the pulsating stars that have been studied for a very long time.  However many questions regarding their pulsation and damping mechanisms still exist.

In stellar physics open questions relating to AF stars concern:
\begin{itemize}
\item the impact of rotation on the stellar structure and its evolution,
\item the understanding of magnetic braking, the loss of a star's angular momentum through magnetic fields,
\item physical processes in the interior of stars which causes pulsation instability,
\item the high precision determination of \teff\ from the analysis of the highly-temperature sensitive H lines.

\end{itemize}

GDR3 contributes to these open questions in many aspects.  
Firstly, the time series for a total of about 10 million variable stars, of which a large fraction are A--F type, are available in the archive.   This variability comprises pulsation periods, eclipsing binaries and rotational modulation allowing the determination of rotation period.  
Secondly, the variability catalogue already provides many of these fundamental variability statistics. Some specific examples in the archive are the {\tt gaiadr3.vari\_ms\_oscillator} table which provides frequencies, amplitudes, and phases of the largest detected amplitudes, or the {\tt gaiadr3.vari\_rotational\_modulation} which provides statistics and periods concerning the rotational information due to e.g. spots on stars, or the {\tt gaiadr3.vari\_short\_timescale} table which gives timescales and periods, along with other statistics of sources with short timescale variability in the photometry.

\cite{deridder23} presents a specific exploitation of the variability tables while focussing on pulsations in main sequence stars with M $\ge 1.3$ \msol.    They concluded that the most populated variability class is the $\delta$ Scuti variable class, they identified over 10\,000 $g$-mode pulsators including many $\gamma$-Dor stars (see their Fig.~6),  an analysis of the empirical period-luminosity relation of these stars revealed different regimes which depend on the oscillation period (see their Fig.~12), and they further concluded that stellar rotation attenuates the amplitude of the dominant oscillation mode of these stars.   All of these results come from the analysis of the data available in GDR3.

The astrophysical parameters for A and F stars have been derived by three different Apsis modules.   The first is GSP-Phot, which has been mentioned above, and parameters from this module are available for all sources down to \gmag\ $<19$.  Secondly, some astrophysical parameters have been derived for the cooler A and F stars by GSP-Spec (upper limit for reliable {\tt\_gspspec} is approximately 7\,500 K), but rotational effects are not taken into account in GDR3 in their analysis of the RVS spectra.  The impact of this parameter on the quality of the data can be quite high, and one should use the GSP-Spec flags to quantify this impact, see \href{https://gea.esac.esa.int/archive/documentation/GDR3/Gaia_archive/chap_datamodel/sec_dm_astrophysical_parameter_tables/ssec_dm_astrophysical_parameters.html#astrophysical_parameters-flags_gspspec}{\blue{the online documentation for details}}.    For the cooler F stars, rotation has a smaller impact on the spectral lines, and these parameters are therefore mostly reliable (see flags). 
Finally the ESP-HS module, mentioned in the above section, analyses the A stars by combining, when possible, the BP/RP and the RVS spectra.   As explained earlier, the $v \sin i$ is derived along with the atmospheric parameters, and a comparison between the {\tt vbroad} parameter in {\tt gaiadr3.gaia\_source} that is derived by the RVS processing and the {\tt vsini\_esphs} is shown in Fig.~\ref{fig:apsisHR} right panel for stars with \teff\ between 7\,500 and 10\,000 K.  The one-to-one line is shown to guide the eye, and one can note in general a good agreement between both determinations, although a scatter is seen.  The different values arise from the very different methodologies used to determine the parameter.  For details, one should consult the {\href{https://gea.esac.esa.int/archive/documentation/GDR3/index.html}{\blue{Gaia Archive documentation}}, and the published papers where the processing of these parameters is described \cite{apsis1,fremat23}.

\subsection{Low mass main sequence stars}
Low-mass main sequence stars are of spectral type K, M, L and T.  These are typically faint objects because of their small size and thus low luminosity, and for this reason they are also difficult to detect and characterise.  Below about 0.3 \msol, these stars are fully convective.  Open questions in stellar physics in this low mass regime concern the transition region between fully and partially convective regimes in stellar interiors and the impact on energy transport in the fully convective regime.  The transition region at the stellar -- sub-stellar border is also an area of active research, and this contributes to the understanding of the initial mass function.  As these objects also evolve very slowly -- on the order of 25 to over 100 billion years, many objects are still in their pre-main sequence phase, and indicators of youth or activity contribute to developing an  evolution (pre-main) sequence.  
GDR3 plays an important role for the understanding of these low main sequence objects firstly from their precise photometry and parallaxes, which allows their detection.   Many low-mass companions in binaries have also been detected using their proper motions, and these binaries often have a more massive component which can contribute to determining the low mass component's mass and / or age, e.g. \citep{sarro2023,gold}.  

In addition to the astrometric and photometric information available in {\tt gaiadr3.gaia\_source}, the Apsis system also provides a characterisation of stars in the low mass region.
GSP-Spec provides reliable atmospheric parameters of stars down to a \teff\ of about 3\,600 K, while GSP-Phot also provides these parameters down to 2\,500 K, which is the lower \teff\ limit of the MARCS grid. In addition to these atmospheric parameters, FLAME also provides some evolutionary parameters (luminosity, radius, mass, age) for these same sources, although masses and ages are limited to stars with masses above 0.5 \msol.

An additional Apsis module ESP-UCD ({\it Extended Stellar Parametrizer -- Ultra Cool Dwarfs}) exploits the RP spectra, along with the astrometry and photometry for identification, in order to derive a \teff\ for the very low mass sources, focussing on the range of 500 K -- 2\,700 K.   The module  uses a Gaussian Process regression model that is trained on a set of Gaia RP spectra.  The choice to use this empirical training was due to the shortcomings in the BT Settl spectra, which inhibited a direct use of simulated data.   This latter fact also highlights the important contribution that Gaia DR3 potentially brings to providing constraints on the model spectra of such low mass objects.   In GDR3 the {\tt gaiadr3.astrophysical\_parameters} table one can find the \teff\ for around 90\,000 ultra cool dwarfs, along with a quality flag (0, 1, or 2) with 0 being the most reliable.  Fig.~\ref{fig:distucd} shows a histogram of the UCD \teff, and the colour code indicates the flag (blue = 0, green = 1, grey = 2).  Note the log scale for interpretation of the numbers.   
The inset shows the distribution of these parameters by \gmag\ and parallax.  

Further analysis of these UCDs is presented in \cite{gold}, where the sources with flags = 0 or 1 are coupled with infra-red data from 2MASS \citep{2mass}, and luminosities and radii are derived.   These luminosities and radii are found in the {\tt gaiadr3.gold\_sample\_ucd} table, and one can combine the data from both tables (or more) for further exploration and analysis of these low mass objects.  Fig.~\ref{fig:distucd} right panel illustrates the relationship found between luminosity and \teff. Only error bars for sources with \teff\ $<$ 1\,900 K are shown, and the colour-code shows the logarithm of the $\chi^2$ fit.   One can clearly see a change in the slope of the luminosity -- \teff\ (and radius -- \teff) diagram around 1\,900 K and again at 1\,500 K which could hint at specific interior physical processes taking place which change the output luminosity (and radius).   

Further scientific exploitation of these objects can also be done by examining these objects that are companions to more massive stars, identified through their common motions.   An analysis of their more massive counterpart, such as their ages, can provide important constraints on the nature of these low mass objects and the limits of the stellar / sub-stellar regime, see e.g. Section 5 in \cite{gold}.
The readers are also referred to the lecture notes by \cite{celine} who presents in much more detail the open questions and available data on ultra-cool dwarfs.

\begin{figure}
\includegraphics[width=0.50\textwidth]{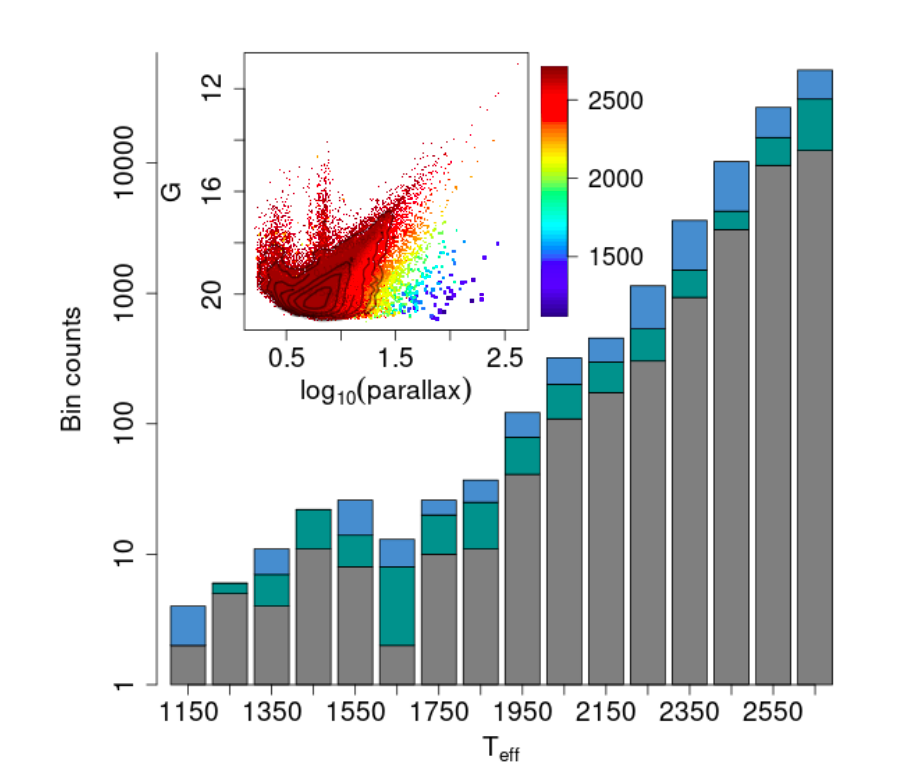}
\includegraphics[width=0.46\textwidth]{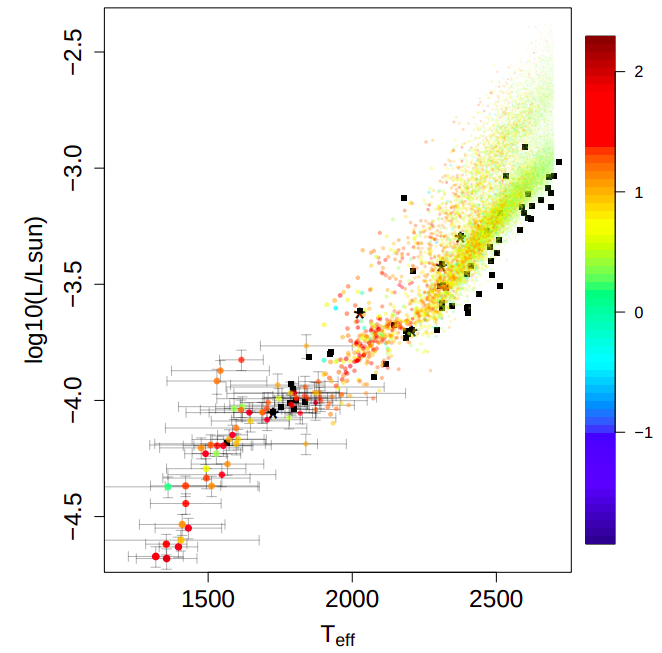}
\caption{{\sl Left:} Distribution of the \teff\ of ultra cool dwarfs in GDR3 produced by the Apsis/ESP-UCD module.  Credit: \creditapsisone. 
{\sl Right:} Luminosities and \teff\ for the UCD sources with flag = 0 or 1.  Error bars are shown only for the sources with \teff\ $< 1\,900$ K.  
Credit: \creditgold.}
\label{fig:distucd}
\end{figure}


\subsection{Exotic stars}
There are many different exotic objects in all regions of the HR diagram, that are not indicated in Fig.~\ref{fig:apsisHR}.   Among these exotic stars, one can mention many 
emission-line stars such as P Cygni stars with winds, accretion inflows in T Tauri stars, rotating discs around Be stars, explosions such as novae and type II supernovae, transfer of material between stars e.g. in symbiotics, and carbon stars, among many others.  Firstly the identification of these objects is of great importance and GDR3 plays an important role here.   Further classification of the objects, and an analysis of their mean and time domain observed properties provides important constraints towards a more concrete understanding of the objects.

As mentioned in Sect.~\ref{sec:hot} the ESP-ELS module provides class probabilities and suggests a class labelling of emission-line stars.  One can revisit the class probabilities to re-label these objects.  There are a total of 57\,511 class probabilities and labels of ELS in GDR3 found in the {\tt gaiadr3.astrophysical\_parameters} table.  

A spectral typing of 218 million sources is also provided.  Apart from the standard O, B, A, F, G, K, and M spectral classes, an additional class of 386\,936 candidate carbon stars 'CSTAR' has been also identified ({\tt spectraltype\_esphs = `CSTAR'}).   Carbon stars are typically asymptotic giant stars with carbon-enriched atmospheres due to either the transfer  of matter in binary systems or to the pollution by nuclear He fusion products from the interior to the exterior layers of stars.  
An analysis of the spectral indices derived from the BP/RP spectra is provided in \cite{gold} where in addition to the class labels,  detailed analysis of the band head strengths of the C$_2$ lines at 482.3 nm and 527.1 nm, and CN lines at 773.3 nm and 895.0 nm of the BP/RP spectra is performed.   An additional sample of 15\,740 stars with the strongest CN features is provided in the {\tt gaiadr3.gold\_sample\_carbon\_stars} table.

An analysis of the H-alpha line in the BP/RP spectra is also provided in GDR3 for 235 million objects. This line is also used for the identification of the ELS: Be stars, Herbig Ae/Be stars, T Tauri stars, active M dwarf stars, Wolf Rayet WC stars and WN stars, and planetary nebula, see the parameter {\tt ew\_espels\_halpha} and related products in the {\tt gaiadr3.astrophysical\_parameters} table.  

In addition to these characterisations, light curves are provided for those stars considered variable in GDR3.     A simple query on the archive such as the following
\begin{verbatim}
select count(*)
from gaiadr3.astrophysical_parameters as ap
inner join gaiadr3.vari_summary as vs on vs.source_id = ap.source_id
where classlabel_espels is not NULL
\end{verbatim}
indicates that there are light curves of 14\,240 sources classified as ELS and also classified in the variability table, and therefore have epoch photometry.   Exploration of these light curves and mean spectra are left to the users.

\subsection{FGK stars\label{sec:fgkstars}}
This section focusses on FGK main sequence, sub-giant and giant stars, which are the most numerous spectral types in GDR3.  They have typical lifetimes on the order of 1 - 15 billion years, so they are ideal sources to trace the star formation history and evolution of our Galaxy.   The determination of their stellar parameters allows one to explore stellar populations and knowledge of their chemical abundances allows one to trace the chemical evolution of the Galaxy, and explore processes of chemical transport in stellar interiors and atmospheres. 

Focussing specifically on stellar physics, the following open questions can be addressed by GDR3: how are chemical elements in stellar interiors transported?  How does magnetism in the interiors and atmospheres of stars govern the stellar structure and evolution?   How is energy transported from the interior to the surface of stars and through their atmospheres? And what is the impact of interface regions such as radiative and convective zones, interiors and atmospheres, and atmospheres and the interstellar medium?
What is the impact of binarity and activity on the evolution of stars?
By focussing on systems where numerous observational constraints can be imposed, such as clusters, binaries, and seismic targets, how well can we determine stellar ages? What is the impact of different micro- and macroscopic processes on the stellar structure, evolution and global parameters and how do these processes influence the knowledge of their ages?
By accurately characterising stars and their ages, what can we learn about evolutionary processes in stars (e.g. rotation), how do exoplanetary systems form and evolve? And how does our Milky Way form and evolve?
Is our own Sun a typical star and how does it compare to other sun-like stars in terms of magnetism, rotation, activity, and chemical composition? And can we place our own Solar System in the context of other planetary systems (ages, masses, radii of host star and its planets)?
All of these questions can be addressed, at least partially, using GDR3.   In the following sections we describe the relevant content of the archive that can allow one to address these questions.

\subsubsection{Atmospheric parameters from BP/RP spectra}
BP and RP spectra for about 200 million sources are available in DR3 for analysis.   As part of the Apsis system, two modules provide estimates of atmospheric parameters and extinction, by exploiting the mean BP/RP spectra, astrometry and photometry.   The first is GSP-Phot, described earlier, which analyses FGK stars for $G<19$ using MARCS (2\,500 K -- 8\,000 K) and PHOENIX (4\,000 K -- 10\,000 K) atmospheric models coupled to evolutionary tracks.   \teff, \logg\, \feh, $A_0$, \ebmr, and distance are derived, along with uncertainties and monte carlo markov chains.  The latter are datalink products and can be downloaded after quering of sources.  
The GSP-Phot products are found in the {\tt gaiadr3.astrophysical\_parameters\_supp} table for all models, with the best library result in the {\tt gaiadr3.astrophysical\_parameters} table and a subset of the parameters are copied to {\tt gaiadr3.gaia\_source}.  One should prioritise the use of the latter two tables for selecting the FGK type stars, and one can then choose to use the homogenous set of MARCS or PHOENIX found in the {\tt supp} table.  
To complement the data in GDR3, a proposed calibration of the metallicities is made available due to a systematic bias observed in the results.  The calibration is made available as a software tool available on the archive webpages, and can be found directly here on github: \href{https://www.cosmos.esa.int/web/gaia/dr3-gspphot-metallicity-calibration}{\blue{https://github.com/mpi-astronomy/gdr3apcal}}.
In addition to the atmospheric parameters, the radius and absolute magnitude are also derived by GSP-Phot.

\begin{figure}
\centering{
\includegraphics[width=0.48\textwidth]{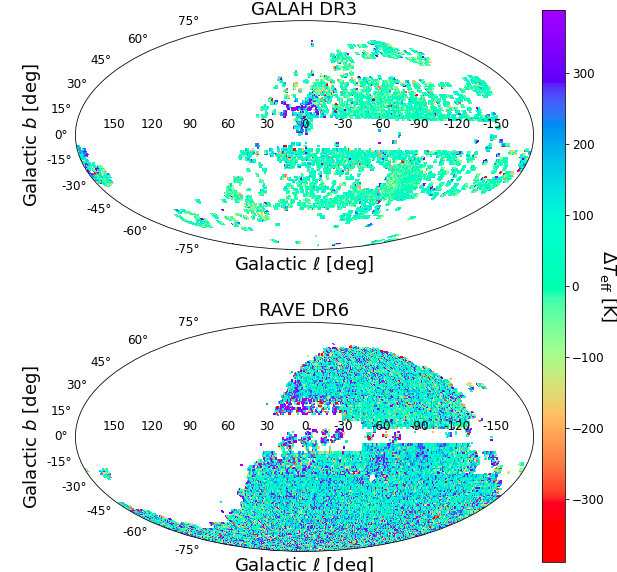}
\includegraphics[width=0.5\textwidth]{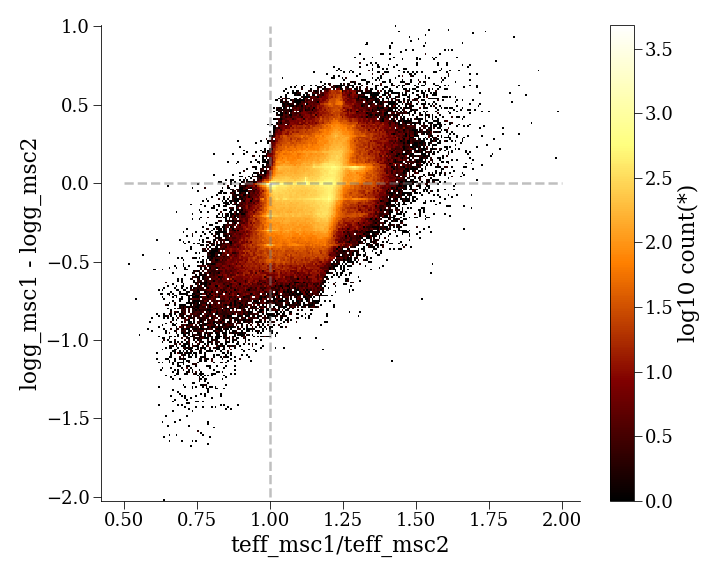}
}
\caption{
{\sl Left:} Differences in \teff\ from GSP-Phot compared to two catalogues from the literature as a function of sky position.  Credit: \creditrene.
{\sl Right:} Distribution of the differences in \logg\ and \teff\ of the two components derived by MSC, with the colour code indicating number of sources.   Credit: \creditapsisone.
}
\label{fig:msc}
\end{figure}

A comparison of \teff\ with external catalogues is shown on the left panel of Fig.~\ref{fig:msc} for GALAH DR3 \citep{galahdr3} and RAVE DR6 \citep{ravedr6}.  Biases in the \teff\ can be seen for regions close to the galactic plane, and this is a result of the high correlation between \teff\ and $A_0$ in the BP/RP spectra.  Making a refined selection on the sources, such as that proposed by \cite{gold}, see below, where astrometric and photometric data are of higher quality, results in much smaller biases when comparing to external catalogues, see e.g. Fig.~{9} from that paper.

A second module also exploits the BP and RP spectra in a similar manner to GSP-Phot, but it assumes that the source is a composition of two individual (non-resolved) components.  It is called the {\it Multiple Stellar Classifier} MSC, and provides for each source two \teff, two \logg, and a unique distance and $A_0$, assuming a solar metallicity prior.   The module is trained on empirical data covering the FGK spectral ranges.   Results for all sources to $G < 18.25$ are provided, and one should use the log-posterior parameter to evaluate how reliable the results are, i.e. how likely the source is to be composed of two non-resolved components.  These parameters are also found in the {\tt gaiadr3.astrophysical\_parameters} table as {\tt teff\_msc1}, {\tt logg\_msc2}, etc. 
Fig.~\ref{fig:msc} right panel illustrates the range of parameter differences that are explored by MSC. Components are generally within 1 dex of \logg\ of each other, while \teff\ ratios vary typically between 0.75 and 1.50.

\subsubsection{Atmospheric parameters from RVS spectra}
The atmospheric parameters \teff, \logg, \feh, and [$\alpha$/Fe] are derived from the RVS spectra by the GSP-Spec module \citep{recioblanco23} for approximately 6 million sources with $G < 15$ focussing on the 3\,500 -- 8\,000 K range, see the accompanying presentation in this school from \cite{alejandra}.  Two algorithms process the data and the results based on the MatisseGauguin algorithm are found in the {\tt gaiadr3.astrophysical\_parameters} table, e.g. {\tt teff\_gspspec}, while the results based on the ANN method are found in the {\tt gaiadr3.\-astrophysical\-\_parameters\-\_supp} table, e.g. {\tt alphafe\_gspspec\_ann}.   
In addition to the atmospheric parameters, a flag containing 42 characters is proposed, with each character being a number 0, 1, 2, or 9 indicating the quality of a source based on a particular parameter, see \href{https://gea.esac.esa.int/archive/documentation/GDR3/Gaia_archive/chap_datamodel/sec_dm_astrophysical_parameter_tables/ssec_dm_astrophysical_parameters.html#astrophysical_parameters-flags_gspspec}{\blue{the online documentation for details}}.  The first 11 characters of the flag are related directly to the quality of the four atmospheric parameters and consider the sources of errors from the input data, e.g. a large {\tt rv\_error}, and the quality of the parametrisation itself.   The other characters refer to the chemical abundances.
A comparison of the \teff, \logg, and \feh\ with the literature for sources with flags 0 or 1 are shown in Fig.~\ref{fig:gspspec}.   A small bias is seen in \logg\ and a calibration is proposed for both $\log g$, [M/H], and the individual abundances.  The description of the calibration is given in \cite{recioblanco23} and can also be found directly on the \href{https://www.cosmos.esa.int/web/gaia/dr3-gspspec-metallicity-logg-calibration}{\blue{archive software tools pages}}.   The reader is referred to the accompanying presentation by \cite{alejandra} for more details including scientific exploitation of the parametrization and the chemical abundances in the context of galactic archaeology.

\begin{figure}
\centering{
\includegraphics[width=0.95\textwidth]{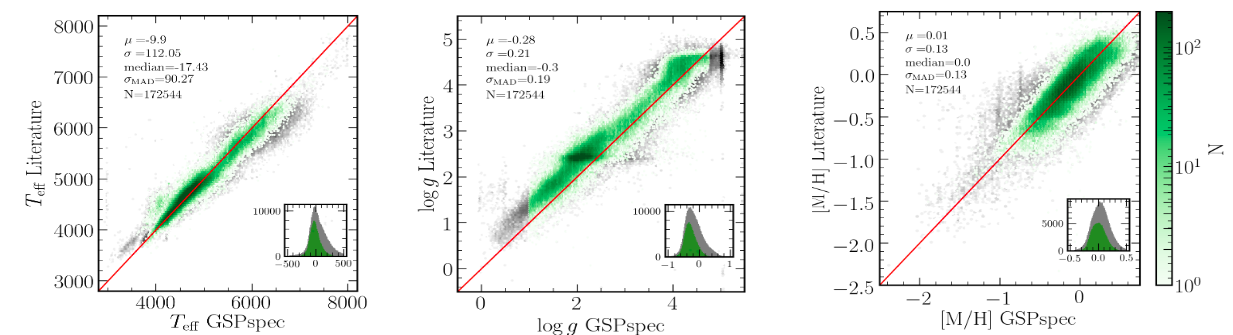}
}
\caption{Comparison to the literature  of \teff, \logg, and \feh\ derived by GSP-Spec from the RVS spectra  for high quality flagged sources.  Credit: \creditarb.}
\label{fig:gspspec}
\end{figure}

\subsubsection{Evolutionary parameters}
Mass and evolutionary parameters are derived by the {\it Final Luminosity Age Mass Estimator} FLAME.   These comprise radius, luminosity, gravitational redshift, masses, ages and evolutionary stage, along with two auxiliary products, a bolometric correction and a flag.
These parameters are produced by using both the GSP-Phot and GSP-Spec atmospheric parameters as input and therefore two sets of evolutionary parameters are available in GDR3.   
The first ones are based on the GSP-Phot parameters and the results for 270 million sources with $G<18.25$ are found in the {\tt gaiadr3.astrophysical\_parameters} table, e.g. {\tt mass\_flame}. 
The second set of results for 6 million sources, based on the GSP-Spec parameters, are found in the {\tt gaiadr3.astrophysical\_parameters\_supp} table, e.g. {\tt age\_flame\_spec}.

\begin{figure}
\centering{
\includegraphics[width=0.48\textwidth]{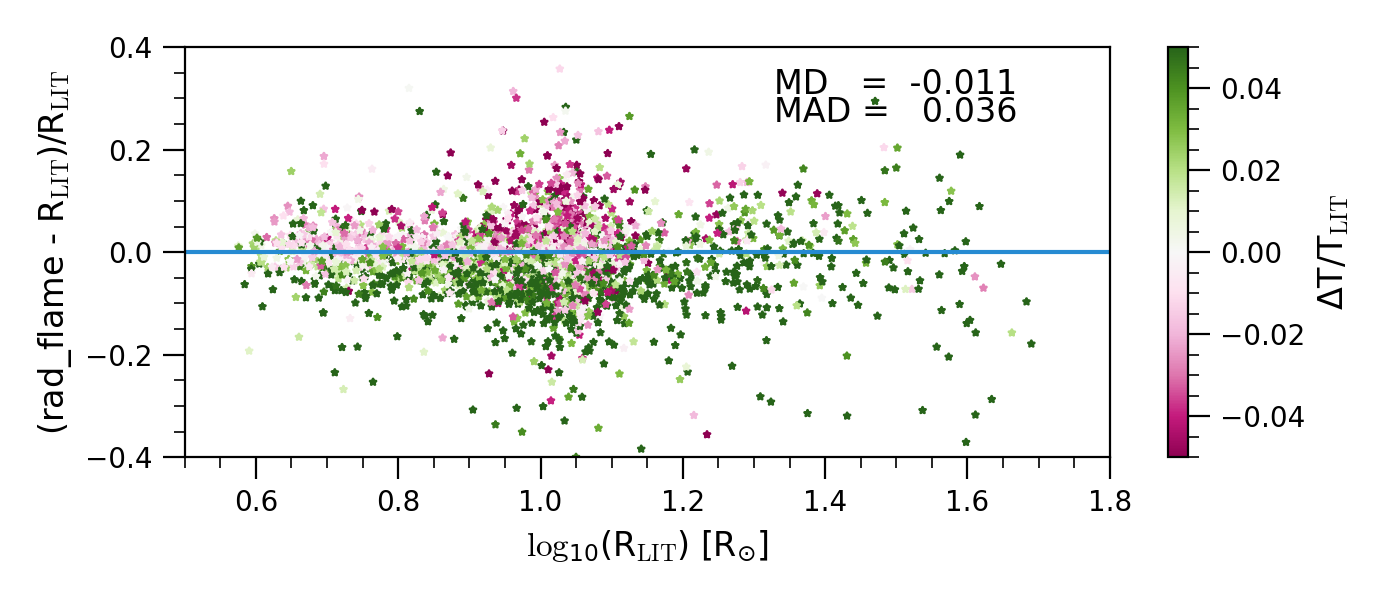}
\includegraphics[width=0.48\textwidth]{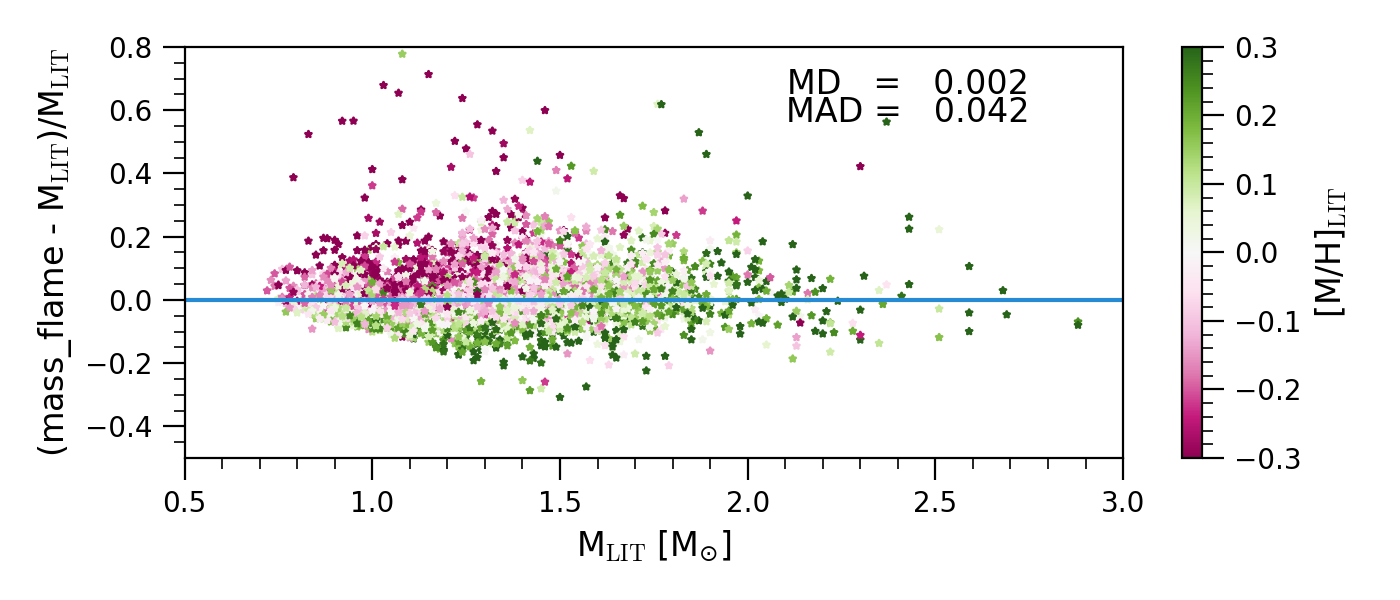}
}
\caption{{\sl Left:} Comparison of {\tt radius\_flame} with asteroseismic radii from \cite{pinsonnealt18} for giant stars.
Colour-code indicates the relative difference in \teff\ estimates assumed in the analysis, which partially explains the dispersion that is found.
{\sl Right:} Comparison of {\tt mass\_flame} with mass estimates from \cite{casagrande11}.   Colour-code indicates the metallicity from the literature.  Credit: \creditapsistwo. }
\label{fig:flame}
\end{figure}

Luminosity, \lum, is derived from the \gmag, parallax (sometimes {\tt distance\_gspphot} is used, see {\tt flags\_flame}), $A_G$, and a bolometric correction based on the input atmospheric parameters.  The bolometric correction is made available to the user and can be accessed from the {\href{https://www.cosmos.esa.int/web/gaia/dr3-software-tools}{\blue{Gaia archive software tools}} or found directly here: \\
\url{https://gitlab.oca.eu/ordenovic/gaiadr3\_bcg}.   Then, using \teff\ and \lum, the radius is derived.  From radius and \logg, the gravitational redshift is then calculated.   Finally, using \teff, \lum, and \feh, the mass, age and evolutionary status are inferred using the BASTI stellar evolution models \citep{basti} with a prior of solar metallicity.  

A comparison with the literature for mass and radius is shown in Fig.~\ref{fig:flame}.  Left and right panels compare giant star radii and main sequence star masses from \cite{pinsonnealt18} and \cite{casagrande11}, respectively.  The colour-code indicates on the left panel the relative difference in assumed \teff\ for the inference, and on the right panel, the literature metallicity, and these can explain some contributions to the scatter that is seen.  These figures illustrate that the results are in general agreement with what is expected.

The impact of the solar metallicity prior has been studied in \cite{LL:OLC-036}, which can be found in \href{https://www.cosmos.esa.int/web/gaia/public-dpac-documents}{\blue{the Gaia DPAC Public Documents}}, and empirical corrections to the mass are proposed given the user's input metallicity, which take the form of \begin{equation}
{M}_{\rm cor} = {M}_{\rm FLAME} - f({M}_{\rm FLAME}, [M/H])
\label{eqn:masscor}
\end{equation} where $f({M}_{\rm FLAME}, {\rm [M/H]})= \sum^{\rm 1}_{\rm i=0} a_{\rm i} x^{\rm i}$ is a linear function to apply to the published mass and the user's \feh\ to produce a corrected mass.  The coefficients are made available in Table 6 of \href{https://dms.cosmos.esa.int/COSMOS/doc_fetch.php?id=1612899}{\blue{\cite{LL:OLC-036}}}, and Fig.~\ref{fig:masscor} left panel illustrates results from simulations showing the residuals, in the sense of $M_{\rm true} - M_{\rm FLAME}$,  when we apply (blue) or not (colours) the proposed corrections for main sequence stars for a metallicity of \feh\ = --0.70.   The right panel shows a comparison of input evolution index assuming \feh\ = --0.70 versus the recovered evolution index assuming a solar metallicity prior for main sequence ($<420$), subgiant (420 -- 490), and giant ($>490$) stars.  While a one to one correspondance is not clear, FLAME still identifies whether the star is main sequence, sub-giant, or giant, even with an incorrect metallicity assumption.   Corrections to the evolutionary status are also proposed in \cite{LL:OLC-036} Table 7.

\begin{figure}
\centering{
\includegraphics[width=0.48\textwidth]{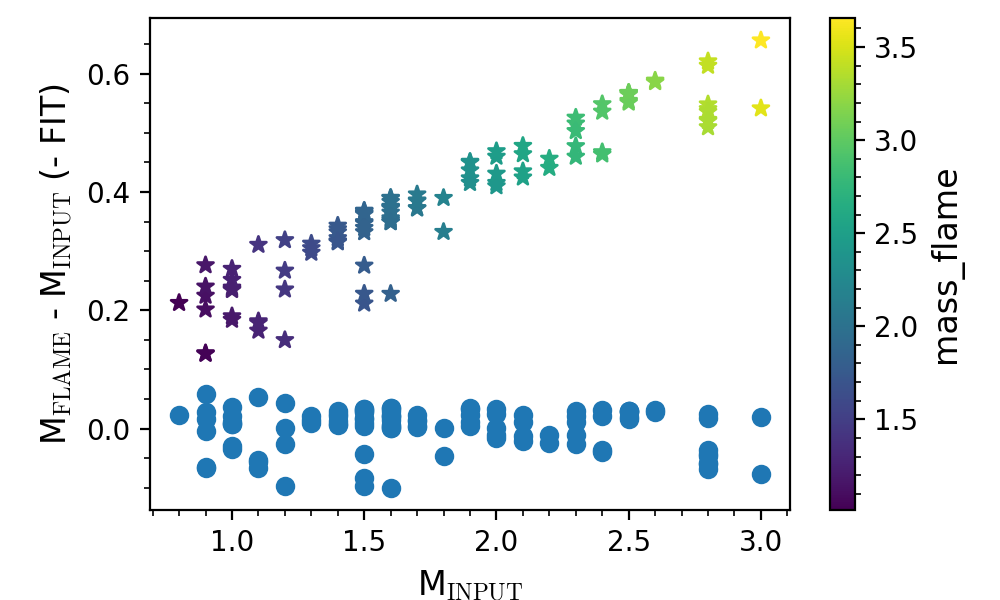}
\includegraphics[width=0.48\textwidth]{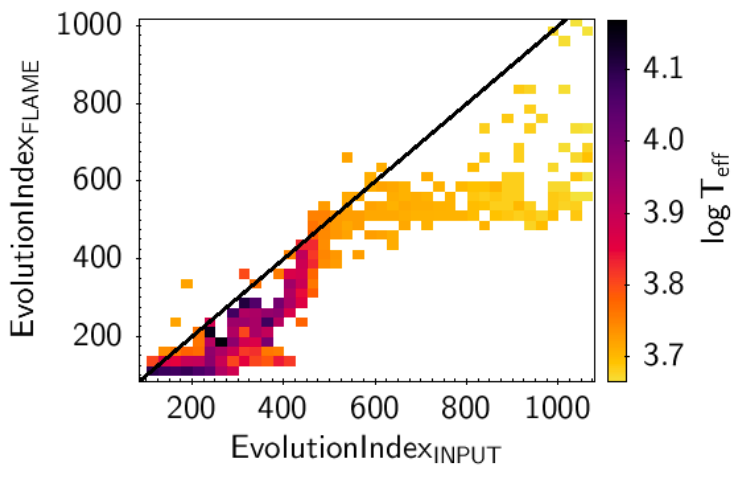}
}
\caption{{\sl Left:} Residuals of mass from FLAME simulations for \feh\ = --0.70, indicating the performance of FLAME using a solar metallicity prior for main sequence stars.  The coloured stars indicate the residuals without applying a correction, while the blue dots show the residuals after applying the empirical corrections.  {\sl Right:}
Comparison of input evolution index assuming \feh\ = --0.70 with the output evolution index assuming a solar prior.  Main sequence, subgiants, and giants are delimited at the values of 420 and 490.  Credit: \href{https://dms.cosmos.esa.int/COSMOS/doc_fetch.php?id=1612899}{\blue{Creevey \& Lebreton, 2022, Public DPAC Documents}}.
}
\label{fig:masscor}
\end{figure}

\subsubsection{Stellar activity}
Within Apsis, the {\it Extended stellar parametrizer -- cool stars}, ESP-CS, analyses the cores of the Ca II IR Triplet lines in the RVS spectra for evidence of stellar activity by comparing the observations with models defining a purely photospheric spectrum.  The excess in the core, given in nm,  is related to activity in the chromosphere of the star or mass accretion, with a value on the order of 0.03 -- 0.05 nm separating the two regimes, see \cite{lanzafame23}.  
In GDR3, about 2 million estimates of {\tt activityindex\_espcs} are derived for stars with $G<15$ and \teff\ between 3\,000 K and 7\,000 K.  An uncertainty and a flag {\tt activityindex\_espcs\_input} indicating the source of the atmospheric parameters (GSP-Phot or GSP-Spec) are also published. Readers are referred to \cite{apsis1,lanzafame23} for full details of the processing and interpretation of the index.

\subsubsection{Metallicities from variability} 
Metallicities and absorption are possible to derive for RR Lyrae stars using pulsation periods and phase differences using multi-colour photometry.  This is described in detail in \cite{clementini23}.  In Gaia DR3 metallicities for 133\,557 RR Lyrae are available in the {\tt gaiadr3.vari\_rrlyrae} table.  A comparison with a subset of metallicities from the GSP-Spec pipeline is shown in Fig.~23 of that paper, which provides an independent validation of the parameters.

\subsubsection{Examples of scientific exploitation}
Several examples of exploitation of the astrophysical parameters for FGK stars are described in \cite{gold}.   Here we describe some of these to give some concrete uses of GDR3.

\paragraph{FGKM gold sample}
A gold sample of about 3.3 million stars is proposed based on very high precision astrometric and photometric data, along with the highly reliable atmospheric and evolutionary parameters from Apsis.   These data can be found in the archive in the table {\tt gaiadr3.gold\_sample\_fgkm\_stars}. The distribution of these sources on the sky is shown on the left panel of Fig.~\ref{fig:golddist}.   This is of interest to examine because it is mostly dominated by the parallax precision and signal-to-noise ratio which is a result of the number of observations for each source and their brightness.   One can immediately interpret that there is not a homogenous coverage in terms of quality of data across the full sky, and one must invoke some quality indicators for scientific exploration, e.g. uncertainties, flags, auxiliary information, such as {\tt ruwe}.   This work makes a very highly selective cut on many parameters in terms of quality because the aim was to produce a relatively small sample with the best data quality.   
The resulting parameters for the final sample is shown in the HR diagram on the right panel.   Plotting a similar HR diagram, but colour-coding by the stellar age, reveals a gradual expected gradient which indicates the high precision of these ages, see their Fig.~12.  

\begin{figure}
\centering{
\includegraphics[width=0.48\textwidth]{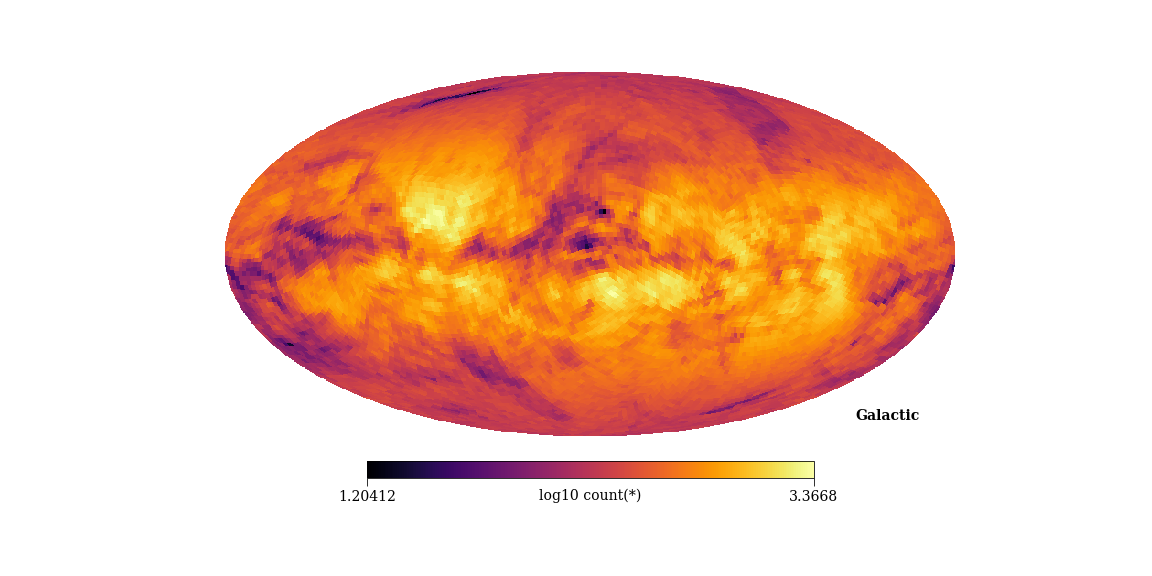}
\includegraphics[width=0.45\textwidth]{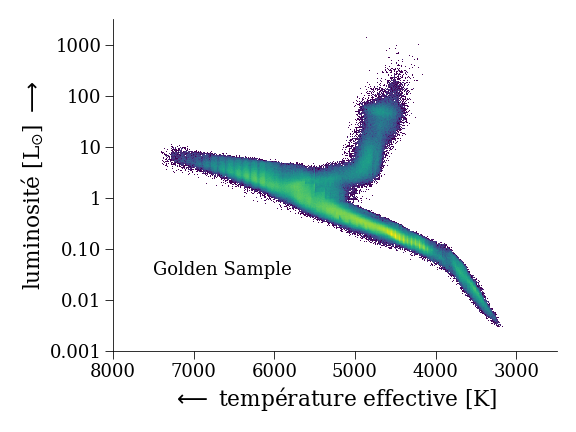}
}
\caption{{\sl Left:}
Distribution of sources on the sky for the FGKM gold sample comprising approximately 4 million stars. 
{\sl Right:}
Resulting HR diagram using the gold sample with colour-code indicating number density of stars. Credits: adapted from \creditgold.
}
\label{fig:golddist}
\end{figure}

\paragraph{Exoplanetary stellar properties}
Cross-matching the golden sample of FGKM stars with known exoplanet host stars using \url{http://exoplanets.org/} dated March 2022, results in a sample of about 600 planets. About 100 of these have measured radial velocities and light curve transits from the literature, allowing the derivation of masses and star-planet relative radii.   We used the stellar masses and radii from the gold sample in order to derive absolute radii, masses, and ages of the 96 exoplanetary systems.   The planet properties are made available through CDS at 
\href{https://cdsarc.cds.unistra.fr/viz-bin/cat/J/A+A/674/A39}{\blue{\tt J/A+A/674/A39}} \citep{goldcat}, and Fig.~\ref{fig:planets} illutrates planet mass -- radius diagrams for Earth-like (left) and Jupiter-like (right) planets, along with several models of planet interiors from \cite{guillot15} and \cite{zeng16}.   

\begin{figure}
\centering{
\includegraphics[width=0.48\textwidth]{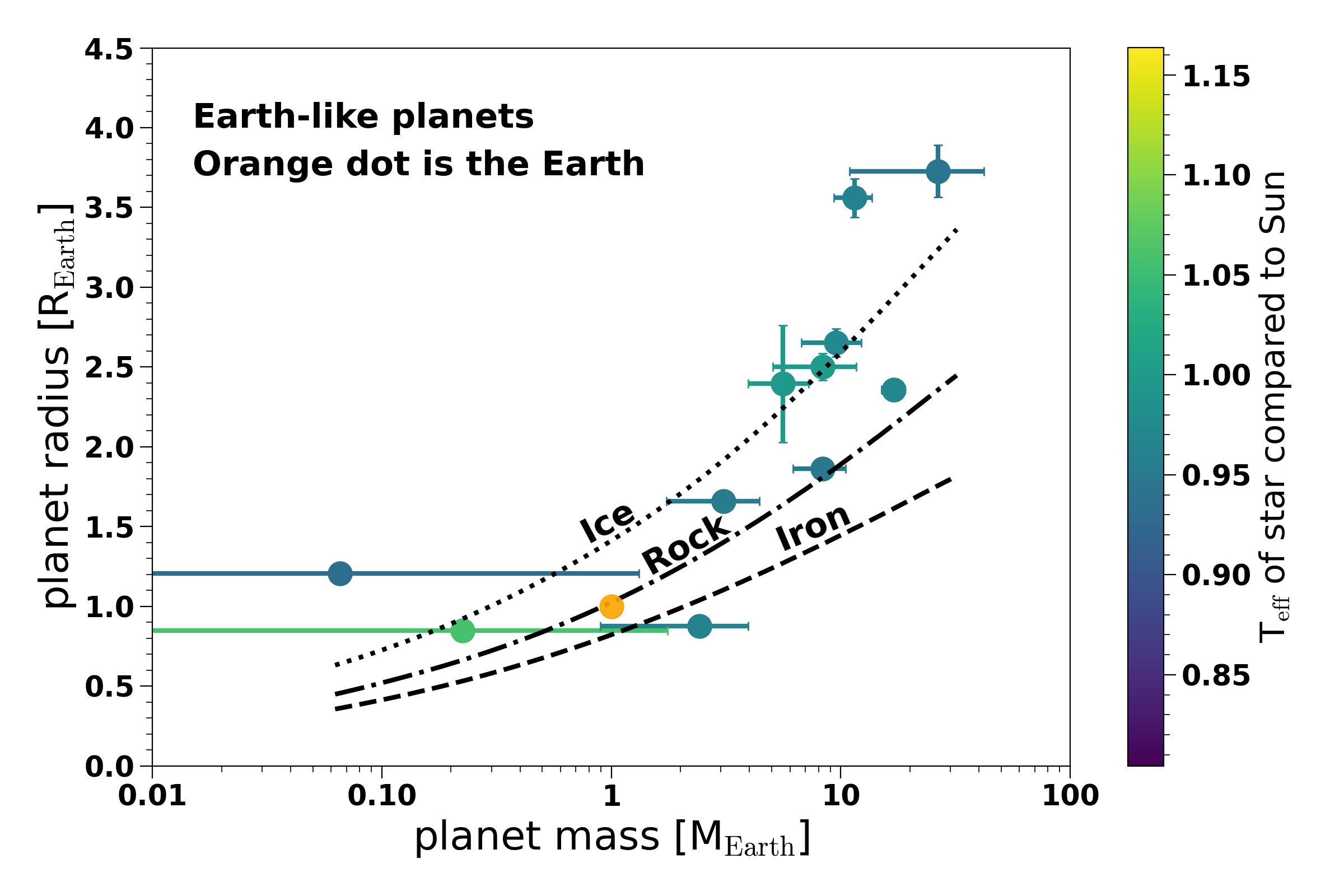}
\includegraphics[width=0.48\textwidth]{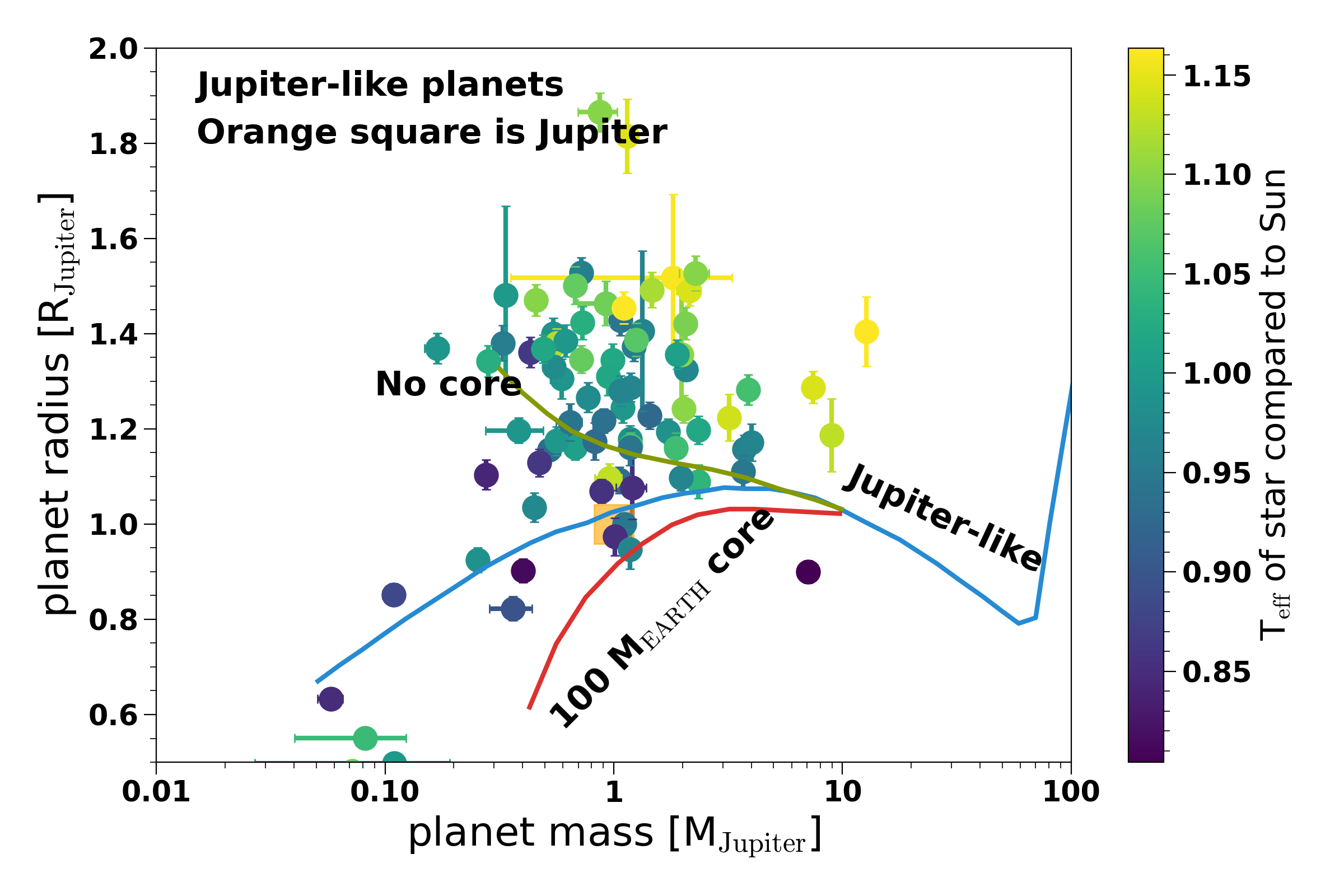}
}
\caption{Planet mass radius diagrams for Earth-like (left) and Jupiter-like (right) planets.  The solar system planets are shown in orange, and several theoretical model predictions are overlaid indicating the stellar bulk composition and density.  Credit: adapted from \creditgold. } 
\label{fig:planets}
\end{figure}

\paragraph{Solar analogues}
Our Sun is the main reference point for understanding the structure and evolution of stars and their atmospheres.  However as it can not be observed with the same instruments as those used for other stars, it is not possible to put it on the same scale with absolute accuracy.   We therefore rely on solar analogues which are stars that resemble the Sun in terms of specific parameters, which depend on the astrophysical question in hand.  
For example, in order to derive accurate \teff\ and \feh\ of other stars, we would aim to choose stars with the same \logg, \teff, and \feh\ of the Sun, while   
if we aim to study age-chemical abundance relations, then we would select stars that also have similar masses to the Sun.

Finding solar analogues has not been a very easy task, as it relies on obtaining many different types of measurements of several candidate analogues, and observing, reducing and analysing these observations.   With Gaia DR3 providing a homogenous catalogue of stellar parameters for the full sky, this job becomes a lot easier.   

A selection of 5\,863 solar analogues was made based on matching the Sun's atmospheric parameters, and mass and radius, to the parameters provided by GSP-Spec and FLAME, along with high precision parallaxes.   Of these, 1046 have published RVS spectra.   In Fig.~\ref{fig:rvssolar} we show the the median RVS spectra of the 1046 solar analogues and the {\it scatter} includes respectively 68\% and 90\% of these analogues.   The similarity of the spectra attests to the correct parametrization in Gaia DR3.   
The full sample of 5\,863 analogues can be found on the archive table {\tt gaiadr3.gold\_sample\_solar\_analogues}.   

Performing the same research using the GSP-Phot and FLAME parameters results in a total of 234\,779 candidates, 7884 of which have RVS spectra and all of which have XP spectra.   We note, however, that as XP spectra are affected by interstellar extinction, one must consider the scientific case before selecting the sources.   

By using the {\tt gaiadr3.gold\_sample\_solar\_analogues} tables and choosing those sources with very low extinction, by selecting on {\tt azero\_gspphot} $< 0.001$, we obtain 682 solar analogue sources for which we can derive the absolute magnitude and the intrinsic colours of the Sun, see \cite{gold}.  
Using the photometric data directly from the Gaia archive, we obtain
\begin{equation}
(G_{BP} - G_{RP})_{\odot}  =  (0.818 \pm 0.029)~ \rm{mag}
\end{equation}
\begin{equation}
(G_{BP} - G)_{\odot}  =  (0.324 \pm 0.016)~\rm{mag}
\end{equation}
\begin{equation}
(G - G_{RP})_{\odot} = (0.494 \pm 0.020)~\rm{mag}
\end{equation}
By combining the resulting sources with the data from \cite{montegriffo23b} using the \\
{\tt gaiadr3.synthetic\_photometry\_gspc} table of synthetic magnitudes derived from the XP spectra, 
we can further calculate many other solar colours, e.g. $(r-i)_{\rm SDSS,\odot} = (0.134 \pm\ 0.012)$ mag, 
$(i-z)_{\rm SDSS,\odot} = (0.33 \pm\ 0.011)$ mag, or 
$(B-V)_{\odot}=(0.64 \pm 0.03)$ mag.  The example query to extract the data from the three tables is given as follows
\begin{verbatim}
select gs.source_id, b_jkc_mag, v_jkc_mag, i_jkc_mag, r_jkc_mag, 
g_sdss_mag, r_sdss_mag, i_sdss_mag, z_sdss_mag, phot_g_mean_mag, 
phot_bp_mean_mag, phot_rp_mean_mag, parallax
from gaiadr3.gold_sample_solar_analogues as gs
inner join gaiadr3.synthetic_photometry_gspc as sp on sp.source_id = gs.source_id
inner join gaiadr3.gaia_source as gg on gg.source_id = gs.source_id
where azero_gspphot < 0.001.
\end{verbatim}
The mean absolute magnitude is $M_{G,\odot} = (4.614 \pm 0.179)$ mag in agreement with the analysis in \cite{apsis1} where $M_{G,\odot} = 4.66$ mag is adopted for the FLAME module, and 
$M_{G,\odot} = 4.67$ mag which was obtained by \cite{casagrande18} using Gaia DR2 data. 

\begin{figure}
\centering
\includegraphics[width=0.7\textwidth]{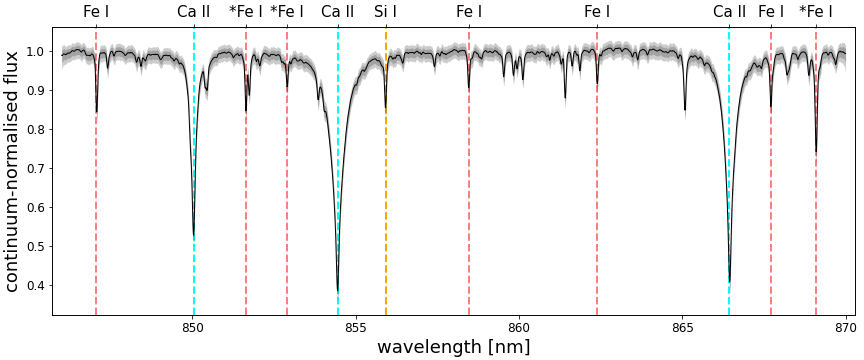}
\caption{The Gaia DR3 RVS spectra of 1046 solar analogues.  The inner and outer grey shaded regions contains 68\% and 90\% of the sample.  Credit: ESA/Gaia/DPAC adapted from \creditgold.}
\label{fig:rvssolar}
\end{figure}

\paragraph{The future of our Sun}
By selecting sources with 1 solar mass and solar metallicity without any constraints on age, we can construct an evolution sequence of a star like the Sun.    This is illustrated in Fig.~\ref{fig:sunlifecycle} where we have selected 5323 stars with the given requirements and plot them on 
a HR diagram in orange over a background of stars from the FGKM gold sample.  In each panel from left to right we choose one star at a particular age and highlight its position, size, and age (the size is not to scale) according to Gaia DR3.   Each highlighted star therefore shows what the Sun would look like at different points in its future, starting with its current age of 4.58 billion years until it reaches the base of red giant branch where it swells up at the age of 11.75 billion years.  
Further detailed studies of these individual stars can be performed to study the fate of our Sun by  refining the age estimates, fundamental, structural, and atmospheric properties, and chemical compositions.  Of particular interest would be the identification of planetary systems around these stars to study the evolution and habitability of future solar systems. 

\begin{figure}
\begin{center}
\includegraphics[width=0.3\textwidth]{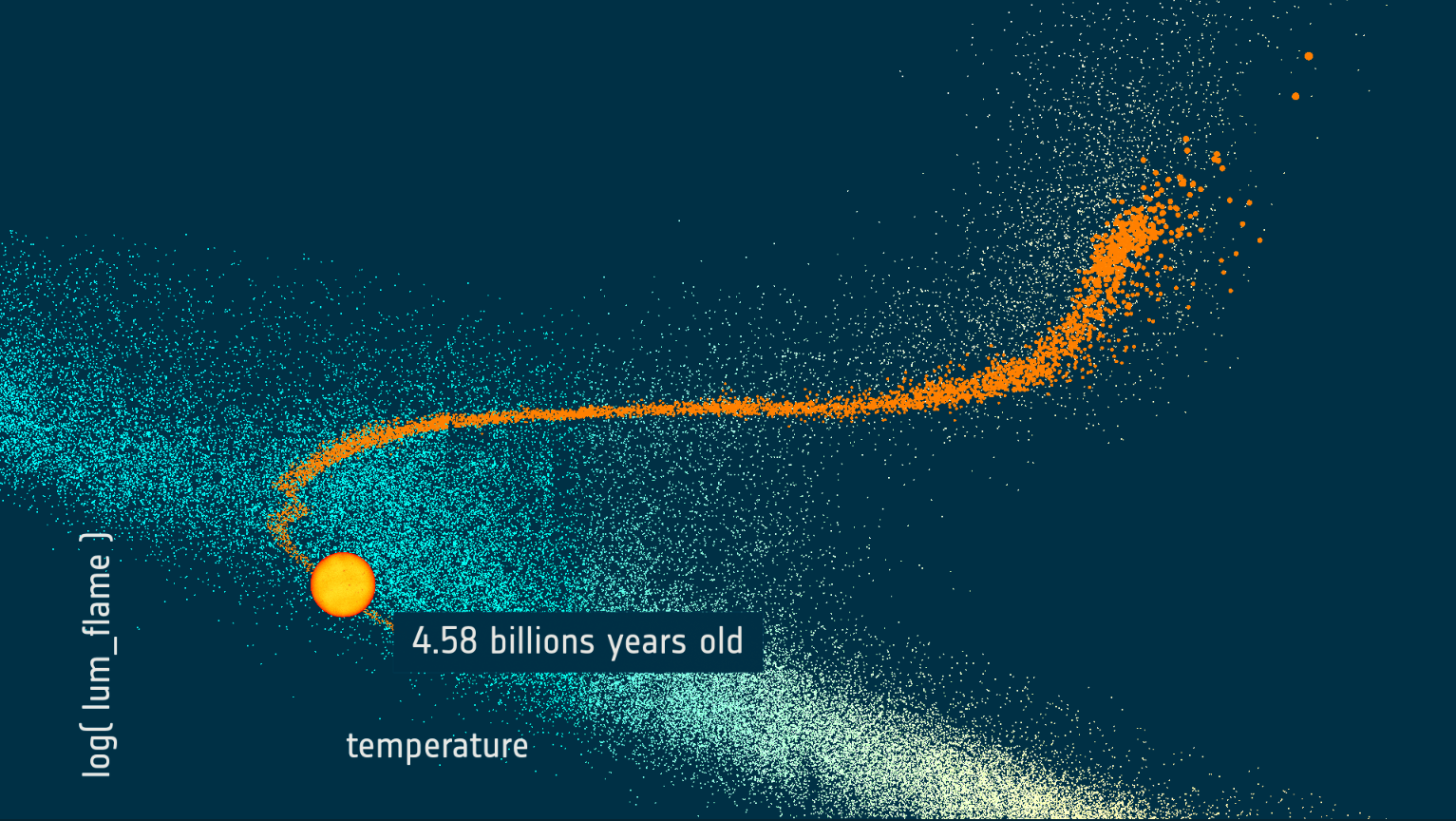}
\includegraphics[width=0.3\textwidth]{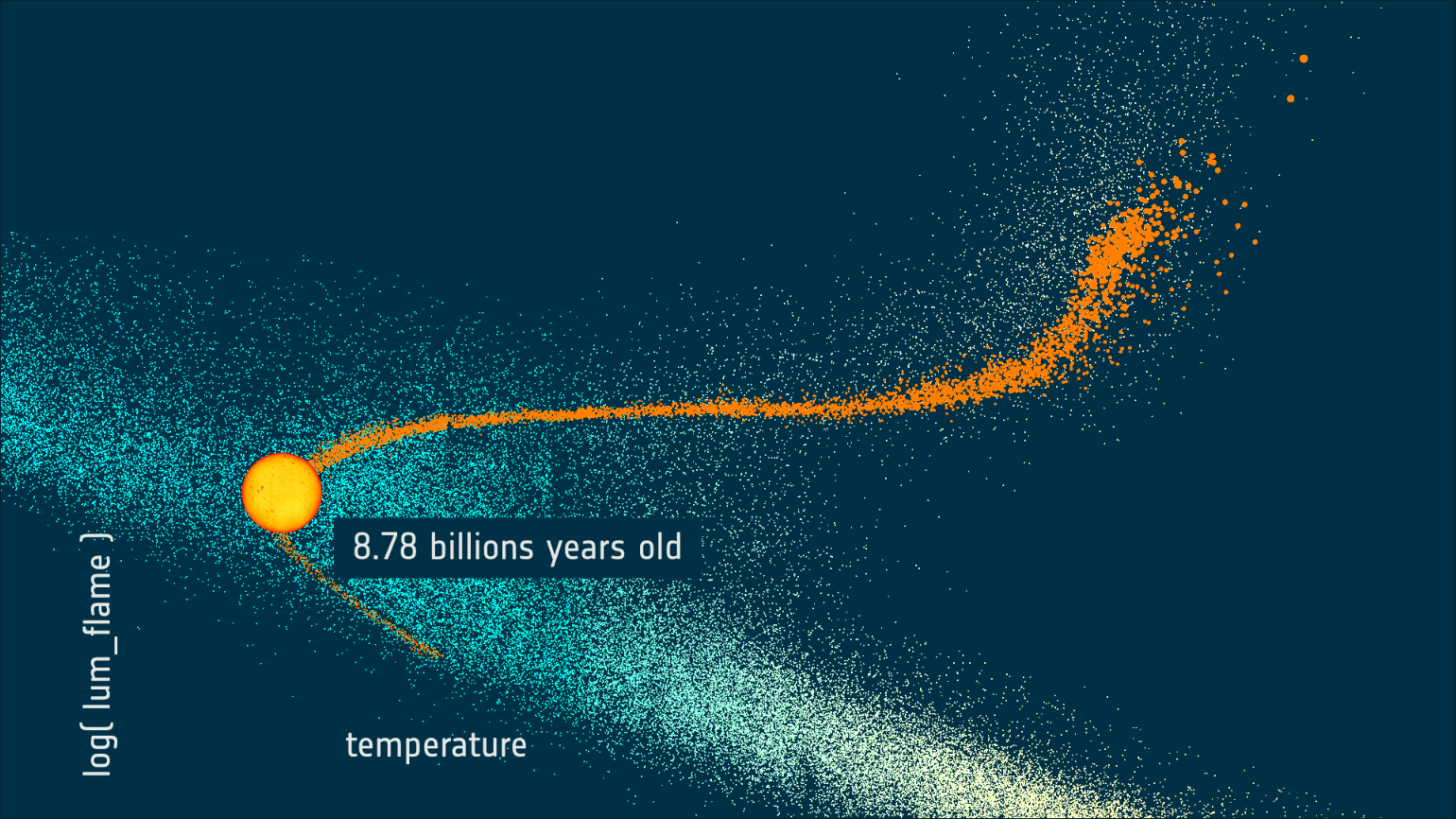}
\includegraphics[width=0.3\textwidth]{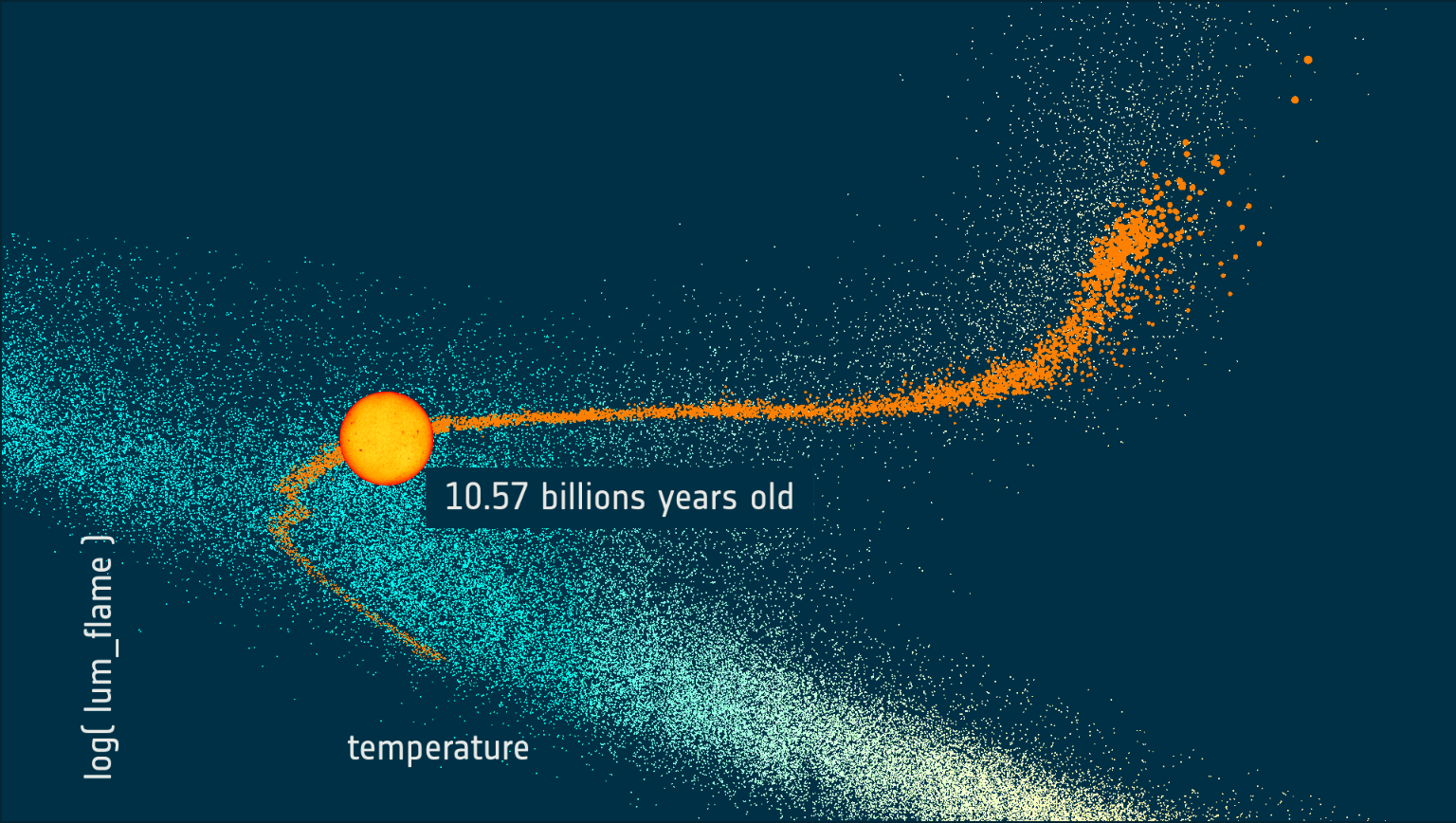}
\includegraphics[width=0.3\textwidth]{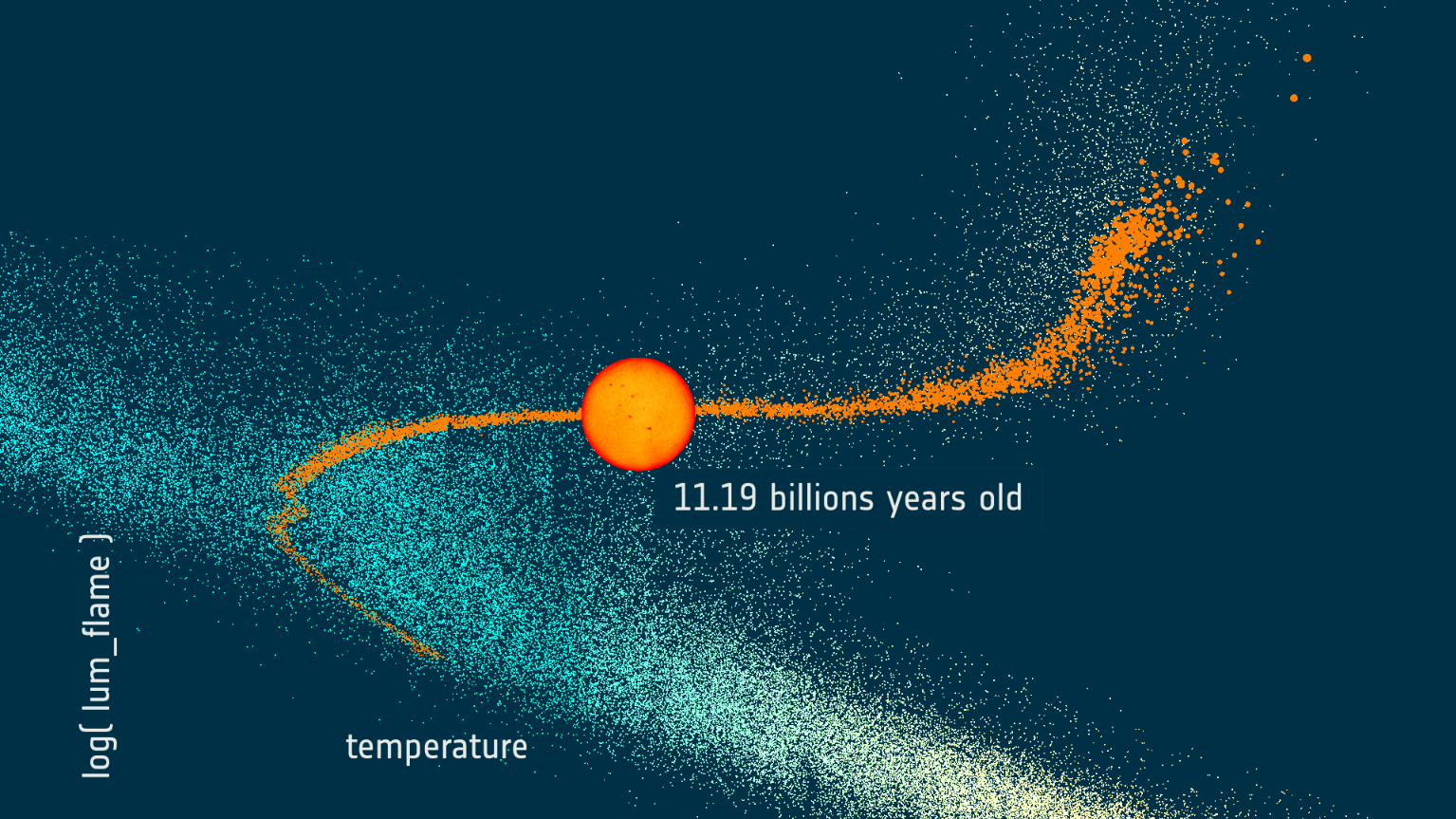}
\includegraphics[width=0.3\textwidth]{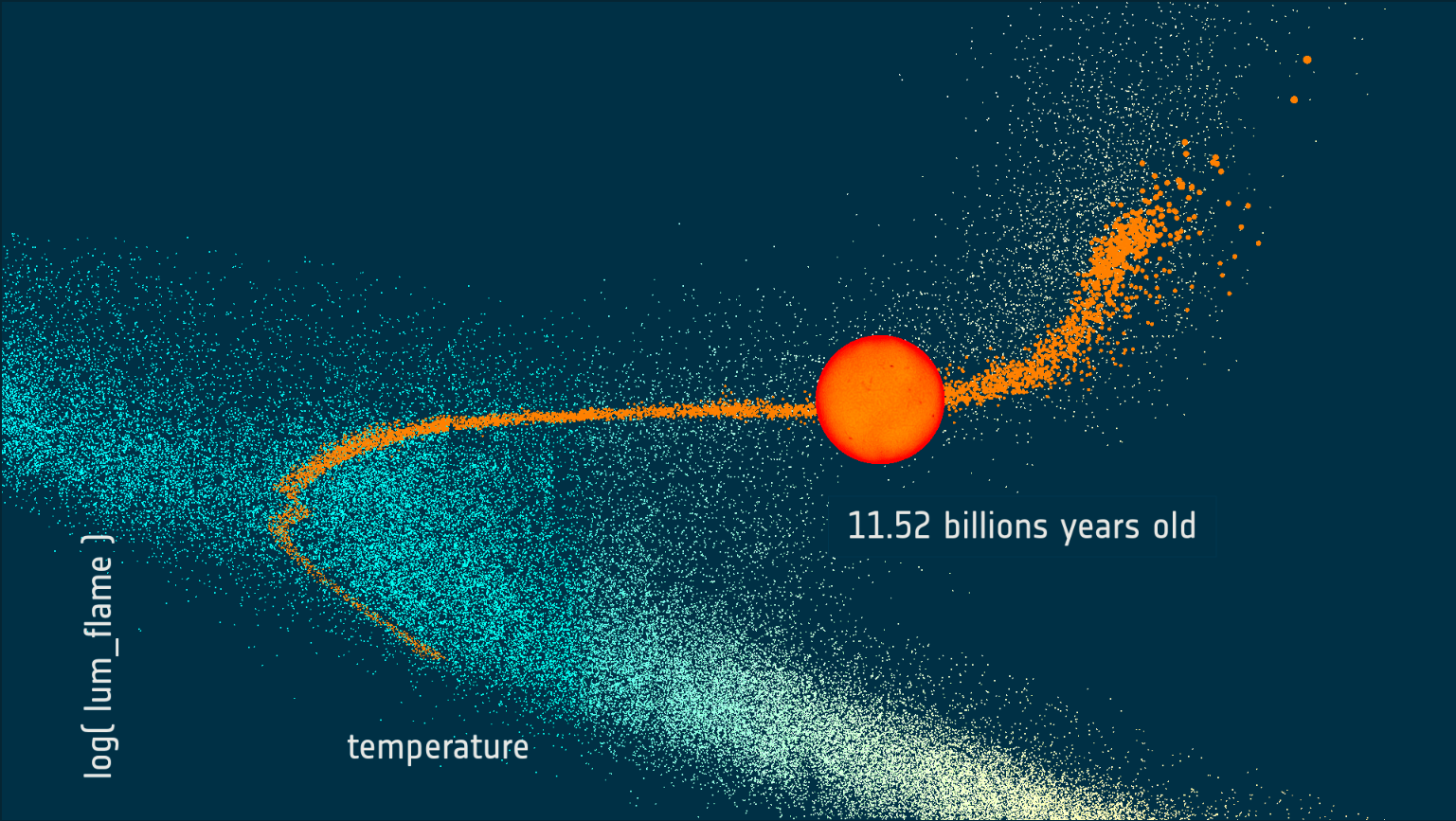}
\includegraphics[width=0.3\textwidth]{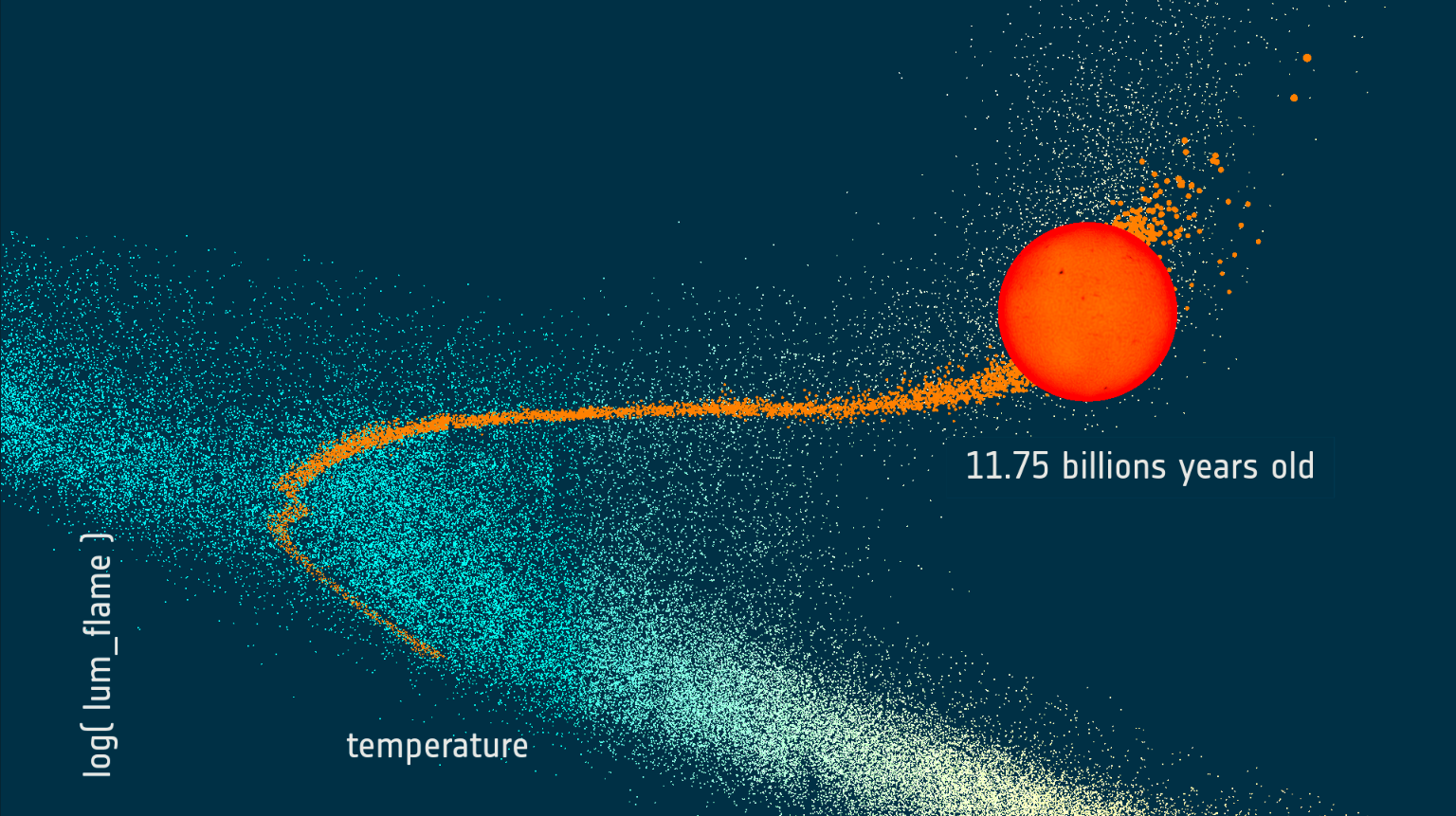}
\end{center}
\caption{HR diagrams showing some of the Golden Sample of FGKM stars.  Overlaid in orange are a selection of stars of one solar mass and solar metallicity from that sample which traces out an evolution sequence.  Each panel highlights one star along the evolution sequence,  with its size (diameter) represented by the size of the symbol, although not to scale, and the age of star indicated in the panel. This represents most of the lifecycle of our own Sun.  Credit: ESA/Gaia/DPAC, adapted from the \href{https://www.youtube.com/watch?v=Nf6kYYtkZpA}{\blue{movie produced by L. Rohrbasser,  K. Nienartowicz, O.L. Creevey, L. Eyer, C. Reyle}}.}
\label{fig:sunlifecycle}
\end{figure}

\subsection{Variability and multiplicity}
Variable stars are those that change their brightness over time, either regularly, semi-regularly, or irregularly.   The changes in brightness can be due to the intrinsic nature of stars (true variable stars), or due to the environment, such as an interaction with another star, or a planet passing in front of the star, or any binary or multiple system.   Variability and multiplicity can also be detected through the astrometric measurements, e.g. photocenter displacements due to huge convective cells in asymptotic giant stars which give rise to a "scattered" position and thus large uncertainties on the astrometric parameters \citep{chiavassa22}, or regular repetitive `circular' patterns due to two or more gravitationally bound objects whose positions are constantly changing, but over a regular time period, see e.g. \href{https://youtu.be/oGqSgBIJtZ0}{\blue{this video illustrating the sky projected motion of a binary star}}.  Both variability and multiplicity can also be detected in radial velocity measurements.    

Two coordination units within Gaia DPAC are dedicated precisely to the study of variability and multiplicity, and they are the specific subjects of other lectures in this school, so details will be left to \cite{laurent,frederic}.
Here we just summarise that all variable objects are found in the Gaia archive in the tables beginning with {\tt vari}, and the {\tt gaiadr3.vari\_summary} table summarises the statistics of the time series of all variable sources along with flags indicating in which variable table the source is found, e.g. source 4071605906863950848 is in both {\tt gaiadr3.vari\_short\_timescale} and {\tt gaiadr3.vari\_eclipsing\_binary}.  The user will then find further details about that source in those tables.
While all of the multiple sources are found in the {\it non-single star} tables, which are those that begin with {\tt gaiadr3.nss}.  Four such tables are available, e.g. the {\tt gaiadr3.nss\_two\_body\_orbit} provides a parallax solution for the two components of binary systems which considers the astrometric motion of both stars in the system.   The astrometry for these sources should, in general, be preferred over the standard astrometric solution given for a single star in {\tt gaiadr3.gaia\_source}.   Users are referred to \cite{laurent} and \cite{frederic} for lectures and examples in this school on both of these topics.

\section{Conclusions}
Gaia DR3 contains a wealth of new data products that can be exploited by the stellar physics community.  Starting with the basic data products described in Sect.~\ref{sec:observables} --- (mean and time-series) photometry, high- and low- resolution spectra, and astrometry --- the user can derive many physical parameters of stars and their environments.   A homogenous analysis of the full Gaia data set is provided in Gaia DR3, based on the time-series and mean data products and this is described in Sect.~\ref{sec:stellarphysics}.  
A general description of the Apsis chain is given, which is responsible for the parametrization of stars and other sources based on the mean data products, while details of the variability and multiplicity analysis is described in \cite{frederic} and \cite{laurent}.  
We then visited different regions of the HR diagram, discussed the open questions, and highlighted how Gaia DR3 can help address these specific questions in stellar physics.  Concrete examples of  scientific exploitation of GDR3 for the specific regions across the HR diagram was also illustrated.  
Example queries to access the data are available in the accompanying tutorials and slides, available at \href{https://ees2023.sciencesconf.org/}{\blue{the website of the Ecole Evry Schatzman 2023}}\footnote{https://ees2023.sciencesconf.org/}.

\begin{acknowledgements}
The author would like to warmly thank Anthony Brown, Francesca De Angeli, and David Katz for providing supporting material which was used to prepare part of this lecture, and 
the organisors of the {\it Ecole Evry Schatzman 2023}, in particular, Carine Babusiaux.   
The author would also like to thank Gaia Coordination Unit 8 for their hard work over many years that resulted in this wonderful catalogue of astrophysical parameters, and to everyone involved in making Gaia possible, from operations, to ground-processing, archive preparation and validation.
This work has made use of data from the European Space Agency (ESA) mission
{\it Gaia} (\url{https://www.cosmos.esa.int/gaia}), processed by the {\it Gaia}
Data Processing and Analysis Consortium (DPAC,
\url{https://www.cosmos.esa.int/web/gaia/dpac/consortium}). Funding for the DPAC
has been provided by national institutions, in particular the institutions
participating in the {\it Gaia} Multilateral Agreement.
\end{acknowledgements}

\bibliographystyle{aa}  
\bibliography{ees2023-creevey} 

 \end{document}